\documentclass[a4paper,10pt,oneside]{article}
\usepackage[latin1]{inputenc}
\usepackage{amsmath}
\usepackage{amsfonts}
\usepackage{amssymb}
\usepackage{graphicx}
\usepackage{footnote}
\usepackage{float}
\usepackage{subcaption}
\DeclareGraphicsExtensions{.bmp,.png,.pdf,.jpg,.eps}
\usepackage{physics}
\usepackage{mathrsfs}
\usepackage{cite}
\usepackage[colorlinks,urlcolor=blue,citecolor=blue]{hyperref}
\advance \textheight by \topskip
\usepackage{amsmath}
\usepackage[english]{babel}
\makeatletter
 \@addtoreset{equation}{section}
\makeatother
\usepackage[latin1]{inputenc}
\usepackage{amsmath}
\numberwithin{equation}{section}

\advance \textheight by \topskip
\usepackage{amsmath}
\usepackage[english]{babel}
\DeclareMathAlphabet\mathbfcal{OMS}{cmsy}{b}{n}
\DeclareMathAlphabet{\boldmathe}{T1}{cmr}{bx}{it}
\newcommand{\mbf}[1]{\boldmathe{#1}}

\def\vr{\mbf{r}}
\def\vs{\mbf{s}}

\def\be{\begin{equation}}
\def\ee{\end{equation}}

\def\R{\mathbb R}

\def\C{\mathbb C}

\def\be{\begin{equation}}
\def\ee{\end{equation}}

\def\R{\mathbb R}

\def\C{\mathbb C}

\def\vr{\mbf{r}}

\def\vs{\mbf{s}}

\def\be{\begin{equation}}
\def\ee{\end{equation}}

\def\R{\mathbb R}

\def\C{\mathbb C}

\def\Ti{\text{i}}

\usepackage{todonotes}

\usepackage{orcidlink}

\begin{document}
%%%%%%%%%%%%%%%%%%%%%%%%%%%%%

\begin{center}
{\LARGE \bf
The soliton nature of the  super-Klein tunneling effect\\ 
}
\vspace{6mm}
{\Large Francisco Correa \orcidlink{0000-0003-1735-2822}$^a$, Luis Inzunza \orcidlink{0000-0002-3659-6956}$^{a}$ and 
Olaf Lechtenfeld \orcidlink{0000-0003-1947-484X}$^b$
}
\\[6mm]

\noindent ${}^a${\em Departamento de F\'isica\\ Universidad de Santiago de Chile,  Av. Victor Jara 3493, Santiago, Chile}\\[3mm]
\noindent ${}^b${\em
	Institut f\"ur Theoretische Physik and Riemann Center for Geometry and Physics Leibniz Universit\"at Hannover
	Appelstra\ss e 2, 30167 Hannover, Germany}
\vspace{12mm}
\end{center}

\begin{abstract}
We establish a relationship between the Davey--Stewartson II (DS II) integrable system in $(2{+}1)$ dimensions and quasi-exactly solvable planar interacting Dirac Hamiltonians that exhibit the super-Klein tunneling (SKT) effect. The Dirac interactions are constructed from the real and imaginary parts of breather solutions of the DS II system. In this framework, the SKT effect arises when the energy is tuned to match the constant background of the soliton, while the resulting Dirac Hamiltonians simultaneously support bound states embedded in the continuum. By imposing the SKT boundary conditions, we employ Darboux transformations to construct a general three-parameter family of DS II breather solutions that can be mapped to Dirac Hamiltonians. At the initial soliton time, the corresponding Dirac systems form a massless two-parameter family of Hermitian models with nontrivial electrostatic potentials. As the soliton time evolves, the systems become $\mathcal{PT}$-symmetric and develop a nontrivial imaginary mass term. Finally, when the soliton time is taken to be imaginary, the construction yields Hermitian Dirac systems that lack time-reversal symmetry. In all cases, we identify the emergence of quasi-symmetry transformations that preserve the SKT subspace of states while not commuting with the full Hamiltonian.
\end{abstract}

\section{Introduction}
Solitons are typically understood as field or wave configurations that constitute exact solutions of integrable systems with infinitely many degrees of freedom, where dispersion and nonlinearity exactly balance each other \cite{Soliton}. Although the formal definition of integrability often requires a more sophisticated mathematical framework, such systems share a number of well-known structural properties. In particular, they possess infinitely many conserved quantities in involution, which in turn define a hierarchy of related equations connected through the Lax pair formalism \cite{Lax} or, equivalently, through the zero-curvature representation \cite{ZS}. In many cases, explicit solutions can be constructed by means of the inverse scattering transform \cite{Fokas}, or through B\"acklund and Darboux transformations \cite{Matveev}. The relevance of solitons is now indisputable, as they appear in diverse branches of physics, including nonlinear optics \cite{optics}, gravitational models \cite{gravsol}, condensed matter systems \cite{cmss}, among many others.

In one-dimensional quantum mechanics, it is well known that reflectionless systems exhibit a distinct ``soliton nature". A reflectionless potential is characterized by the absence of backscattering, its reflection coefficient vanishes for all energies, a concept first introduced by Kay and Moses \cite{KM}. In their seminal work \cite{Gardner}, Gardner et al. established a remarkable link between the solitonic solutions of the Korteweg--de Vries (KdV) equation (a paradigmatic example of an integrable system) and the potentials of the stationary Schr\"odinger equation. Through the inverse scattering method, soliton solutions can be constructed directly from scattering data by imposing the condition of zero reflection in the continuous spectrum. For instance, the one-soliton solution corresponds to the well-known P\"oschl-Teller potential \cite{Fluge, hidden, amgp}, whose spatial profile evolves in time according to the soliton dynamics.

This correspondence arises because the Schr\"odinger Hamiltonian can be identified as one of the operators in the Lax pair formulation of the KdV equation \cite{Lax}, where the potential function itself evolves according to the KdV dynamics. Within this framework, many solution-generating techniques in quantum mechanics admit a dual interpretation in soliton theory. In particular, the Darboux (or supersymmetric) transformation \cite{Matveev}, which generates ``supersymmetric partner" Hamiltonians from a given one, can be used to construct new quantum models with controlled spectral properties. Since the Lax pair formulation is covariant under such transformations, the supersymmetric partners of KdV potentials are themselves solutions of the KdV equation. As a consequence, this procedure preserves the scattering data, allowing for the systematic generation of new reflectionless potentials with additional bound states. Starting from the free particle, which is manifestly reflectionless, one can construct nontrivial potentials through successive Darboux transformations. These families of potentials correspond precisely to the 
$n$-soliton solutions of the KdV hierarchy, which explains why the P\"oschl--Teller potential naturally emerges as the standard one-soliton solution. The notion of reflectionless potentials has also been extended to time-dependent Schr\"odinger Hamiltonians, where they are related to soliton solutions of multicomponent nonlinear Schr\"odinger equations \cite{nogami}.

In the relativistic one-dimensional setting, the connection between reflectionless potentials and integrable models also persists. Purely transmitting static potentials in the one-dimensional Dirac equation have been shown to be related to integrable equations such as the modified KdV and the nonlinear Schr\"odinger equations, as demonstrated in a series of works for Lorentz-scalar \cite{relkm}, pseudoscalar \cite{pseudo}, and mixed scalar-pseudoscalar potentials \cite{ttn, tn1, tn2}. The connection with integrable systems and soliton solutions is, once again, not accidental. These classes of potentials are associated with self-consistent, static solutions of the Gross--Neveu and Nambu--Jona--Lasinio models \cite{gn,dhn,joshua,shei,ttn, tn1, tn2}, and they can also be constructed via Darboux transformations \cite{ap, twisted}. The generalization to time-dependent relativistic transparent potentials was developed in \cite{dt1, dt2, dt3}. These solutions are likewise connected to exactly solvable fermionic quantum field theories and, once more, to integrability, now through the sinh-Gordon equation.

Although the connection between reflectionless potentials and integrability is well established in one-dimensional systems, total transmission can also occur in more complex relativistic and higher-dimensional settings. This naturally raises the question of whether an underlying integrable structure is present in these cases as well. In this article, we address this issue by focusing on the so-called super-Klein tunneling (SKT) effect. The phenomenon originally referred to as the ``Klein paradox" was predicted within Dirac theory \cite{KleinTun} and indicates that electrostatic potentials can become nearly transparent to electron transmission at specific energies and angles of incidence. This effect, now commonly known as Klein tunneling, was later predicted \cite{KleinGr} and experimentally observed \cite{KleinExp} in Dirac materials, whose low-energy excitations are governed by Dirac-type equations. Super-Klein tunneling represents an even more striking scenario, characterized by perfect transmission that is independent of the angle of incidence. This effect was predicted for spin-1 systems \cite{SKT1,SKT2}, later for spin-$\frac{1}{2}$  materials \cite{asym1}, and more recently for Klein--Gordon particles \cite{SKTKlein}. A comprehensive review of these developments can be found in \cite{bomj}.

To address this problem, we require three key ingredients: (i) a $(2+1)$-dimensional integrable system admitting a Lax pair formulation, from which a planar Dirac equation can be identified; (ii) a well-established Darboux transformation procedure; and (iii) the imposition of boundary conditions compatible with SKT. Within this framework, we consider the defocusing Davey--Stewartson II (DS II) equations \cite{DS}. In the corresponding Lax pair formulation \cite{LaxDS}, one can identify a nontrivial planar Dirac equation at a specific energy eigenvalue, up to a global Pauli matrix factor.  In this construction, a scalar potential arises from the real part of the soliton solution, with an imaginary, but $\mathcal{PT}$-symmetric, mass term\footnote{$\mathcal{PT}$-symmetric systems are non-Hermitian models invariant under combined parity and time-reversal transformations\cite{PT1,PT2}. Despite their non-Hermiticity, they can possess real spectra and a consistent unitary time evolution, and they have also been shown to exhibit reflectionless and integrable features \cite{cjp}.}. The presence of the Pauli matrix factor modifies the standard Darboux transformation \cite{Matveev}, leading instead to its asymmetric version of the intertwining relation \cite{asym1,CorInzJak1,CorInzJak2}. Unlike the conventional one-dimensional case, this transformation maps solutions only at a fixed energy eigenvalue. Finally, imposing the SKT boundary conditions reveals that the relevant soliton must be a breather, i.e., a time-periodic solution localized along an infinite cylinder. Consistency further requires that the selected energy level matches the asymptotic behavior of the soliton. Under these conditions, we construct a three-parameter, two-dimensional, $\mathcal{PT}$-symmetric Dirac system exhibiting the SKT effect, where the soliton time enters as one of the potential parameters. At the initial soliton time, the resulting quantum system reduces to a Hermitian, massless, two-parameter model that also displays the SKT effect. For imaginary soliton time, the mass term becomes real and the system remains Hermitian, but time-reversal invariance is lost. Remarkably, all these regimes are governed by an internal quasi-symmetry structure, in the sense that the space of SKT states is invariant under transformations that do not commute with the Hamiltonian, see \cite{Quasi}.

The article is organized as follows. In Section \ref{Sec1}, we present the fundamental aspects of the DS II system, along with its corresponding Darboux transformation. In Section \ref{SecExp}, we construct a specific example of a breather solution for the DS II equation and explicitly relate it to a Dirac Hamiltonian exhibiting the SKT effect. In Section \ref{Sec4}, we identify the appropriate seed solution required to obtain the DS II solution associated with the SKT-Dirac system. This result is then used in Section \ref{SubSecFam} to construct a multi-parametric family of $\mathcal{P T}$-symmetric models. Furthermore, by investigating the underlying supersymmetric structure in \ref{SecSUSY}, we derive the quasi-symmetry generator for the space of states within the SKT regime. Finally, Section \ref{SecDis} is devoted to the discussion and outlook.

\section{ The DS II system}

\label{Sec1}
The Davey--Stewartson (DS) equation describes the evolution of two-dimensional wave packets \cite{DS}. It consists of a coupled system of nonlinear partial differential equations involving one complex-valued field and one real-valued field, and can be regarded as a natural $(2{+}1)$-dimensional generalization of the nonlinear Schr\"odinger equation. The DS system is integrable and admits a Lax pair formulation \cite{LaxDS}, which is covariant under the Darboux transformation \cite{Matveev}. As a consequence, new solutions can be systematically generated from a given one. In this work, we restrict our attention to the defocusing case, also known as the DS II model. The system arises as the integrability condition of the following nonlinear matrix problem,
\begin{eqnarray}
	\label{DS-system}
	h\psi=0\,,\qquad
	M\psi=0\,,\qquad
	h=\partial_y-\Ti \sigma_3\partial_x-U\,,\qquad 
	M=-\partial_\tau+2\Ti\sigma_3\partial_x^2+2 U \partial_x+W\,.
\end{eqnarray}
Here, $\psi$ is a two-component spinor while the ``matrix potentials" $U$ and $W$ are given by 
\begin{align}
&	U=\left(\begin{array}{ll}
		0 & u\\
		-\overline{u} & 0
	\end{array}
	\right)\,,\\
&	W= \left(\begin{array}{ll}
		\frac{1}{2}(w+\Ti Q) &\Ti (\partial_y+\Ti \partial_x)u\\
		\Ti (\partial_y-\Ti \partial_x)\overline{u} & \frac{1}{2}(w-\Ti Q)
	\end{array}
	\right)=\tfrac{1}{2}(w +\Ti \sigma_3Q)+\Ti (\partial_y+\Ti \sigma_3\partial_x)U\sigma_3
	\,,\label{W0}
\end{align}
being $\sigma_i$ with $i=1,2,3$ the Pauli matrices, $u=u(x,y,\tau)$ a complex scalar function, $Q=Q(x,y,\tau)$ a real function, and  $w=w(x,y,\tau)$ an auxiliary real function. The system is invariant under the discrete symmetry,
\begin{equation}
	\label{revTinv}
	\mathcal{T}h\mathcal{T}^{-1}=h\,,\qquad \mathcal{T}M\mathcal{T}^{-1}=M\,,
\end{equation}
where $\mathcal{T}= T\sigma_2$ and $T$ is the complex conjugation operator.
The integrability condition $\partial_\tau\partial_y\psi=\partial_y\partial_\tau\psi$ yields the nonlinear and coupled $(2{+}1)$-dimensional second-order equations 
\begin{eqnarray}
	&\label{DSsys1}
	(\Ti \partial_\tau+\partial_{x}^2-\partial_{y}^2+Q)u=0\,,\qquad
		\partial_y w=- \partial_x(Q+2|u|^2)\,,
		\qquad
		\partial_x w= \partial_y(Q-2|u|^2)\,.
\end{eqnarray}
The auxiliary function $w$ can be eliminated using the integrability condition $\partial_{x}\partial_y w=\partial_y \partial_{x}w$, obtaining as a result 
\begin{equation}
	\label{eqQ}
(\partial_{x}^2+\partial_{y}^2)Q=2(\partial_{x}^2-\partial_{y}^2)|u|^2\,.
\end{equation} 
The first equation in (\ref{DSsys1}) together with (\ref{eqQ}) constitute the DS II equations. The functions $u$ and $S=Q+2|u|^2$ are called the complex amplitude and the real mean flow \cite{DS}, respectively.

The integrability of DS II allows one to construct a new set of solvable operators $(h,M)=(h_1,M_1)$ with  matrix potentials $(U_1,W_1)$ from a given known set  $(h_0,M_0)$ with $(U_0,W_0)$. To this end,  we use the fact that the matrix problem  (\ref{DS-system}) is covariant under a first-order Darboux transformation 
\begin{equation}
	\label{Inter}
	L\,h_0=h_1\,L\,,\qquad
	L\,M_0=M_1\,L\,,
\end{equation}
where the intertwining operator $L$ and the corresponding potentials $U_1$ and $W_1$ are given by
\begin{align}
	\label{LUW}
		&L= \partial_x-\Sigma\,,\qquad 	U_{1}=U_0+\Ti[\sigma_3,\Sigma]\,,\qquad  W_{1}=W_0+2(\partial_y+\Ti\sigma_3\partial_x)\Sigma\,,
		\\	
		&\Sigma=(\partial_{x}\Phi)\Phi^{-1}\,.\label{Sigma}
\end{align} 
In these relations,  $\Phi$ is a  $2\cross 2$ ``seed matrix" satisfying $h_0 \Phi=M_0\Phi=0$. Such matrix is of the form  $\Phi \propto (\phi_1, \phi_2)$, where $\phi_1$ and $\phi_2$ are two known and distinct spinor annihilated by $h_0$ and $M_0$.

In principle, the components of $U_1$ and $W_1$ should define new solutions for the DS II equations. However, for this to occur, they must be invariant under the $\mathcal{T}$ transformation. Therefore, $\Sigma$ must satisfy 
\begin{equation}\label{condition}
	\mathcal{T}\Sigma\mathcal{T}^{-1}=\Sigma\,.
\end{equation} 
without losing generality, this condition can be achieved by simply picking 
\begin{equation}
	\label{condition2}
	\Phi=e^{\frac{\Ti \pi}{4}}(\phi_1,\phi_2)=e^{\frac{\Ti \pi}{4}}(\phi_1,\mathcal{T}\phi_1)=e^{\Ti \frac{\pi}{4}}\left(
	\begin{array}{cc}
	f & \Ti\overline{g}\\
	g& -\Ti \overline{f}
	\end{array}
\right)	\qquad \Rightarrow\qquad  \mathcal{T}\Phi\mathcal{T}^{-1}=\Phi\,,
\end{equation}
where we denoted a particular spinor solution as $\phi_1=(f,g)^T$, with $f=f(x,y,\tau)$ and $g=g(x,y,\tau)$. The global phase factor $e^{\Ti \frac{\pi}{4}}$ in (\ref{condition2}) ensures that the action of $\mathcal{T}$ does not produce an irrelevant phase factor. Its choice is inconsequential for the construction of $\Sigma$ since
\begin{equation}
	\label{SimB}
	\Sigma=(\partial_x\Phi) \Phi^{-1}=(\partial_x (\Phi B)) (\Phi B)^{-1} \, ,
\end{equation} 
for any constant invertible matrix. This ambiguity will be helpful in the following sections. 

Additionally, we have  
\begin{equation}
\label{NonSin}
\text{Det}(\Phi)= |f(x,y)|^2+|g(x,y)|^2>0\,,
\end{equation}
which avoids singularities in $\Phi^{-1}$ and in the new DS II solution. 

With this choice, the new solutions of equations (\ref{DSsys1}) are given by the matrix elements 
\begin{align}
\label{uv}
	u_{1}=(U_{1})_{1,2}\,,\qquad 
	Q_{1}=\Ti[(W_{1})_{2,2}-(W_{1})_{1,1}]\,,\qquad 
	w_{1}=(W_{1})_{2,2}+(W_{1})_{1,1}
	\,.
\end{align}
In the following section, we explore how explicit solutions can be constructed and how they can be related to two-dimensional Dirac operators to nontrivial potentials and mass terms.

\section{The SKT/DS II relation: An example with a bound state in the continuum}
\label{SecExp}
This section is devoted to the construction of a particular solution of the DS II system and to demonstrating how it can be related to a two-dimensional Dirac-type Hamiltonian. We then explicitly construct the scattering states at a specific energy level, chosen to match the background of the DS II solution. We show that these states exhibit total transmission, thereby realizing the SKT effect. In addition, other properties of the system, including bound states and symmetries, are discussed.

Let us start with a constant background DS II solution with matrix potentials
\begin{equation}
	\label{consstantback}
	U_0=-\Ti \lambda \sigma_2\,,\qquad W_0=0\,,
\end{equation}
where $\lambda$ is a nonzero constant value. In such case, the operators $h_0$ and $M_0$ take the form
\begin{eqnarray}
	\label{DS-system 0}
	h_0=\partial_y-\Ti\sigma_3\partial_x+ \Ti \lambda \sigma_2\,,\qquad
	M_0=-\partial_\tau+2\Ti\left(\sigma_3\partial_x^2-\lambda \sigma_2 \partial_x\right)\,.
\end{eqnarray}
By using $\lambda$ as a scale factor,  we introduce the scaled coordinates   
\begin{equation}
	x^\prime =\lambda \,x\,, \qquad 	y^\prime =\lambda\, y\,,\qquad \tau^\prime =\lambda^2\, \tau\,,
\end{equation}
in terms of which the operators now read as 
\begin{equation}
	\lambda^{-1}h_0=\partial_{y^\prime}-\Ti\sigma_3\partial_{x^\prime}+\Ti \sigma_2= h_0^\prime\qquad 
	\lambda^{-2}	M_0=-\partial_{\tau^\prime}+2\Ti\left(\sigma_3\partial_{x^\prime}^2- \sigma_2 \partial_{x^\prime}\right)=  M_0^\prime\,.
\end{equation}
Clearly, working with the scaled coordinates is equivalent to fixing $\lambda=1$ from the beginning, which we assume from now on. 

General solutions of $h_0\psi=M_0\psi=0$ can be formally written as 
\begin{equation}
	\label{generalsol}
	\psi(x,y,\tau)= e^{2\tau\partial_x\partial_y}\psi(x,y)\,,\qquad
	\psi(x,y)=e^{(\Ti\sigma_3\partial_x-\Ti \sigma_2)y}\psi(x)\,,
\end{equation}
where $\psi(x)$ is some arbitrary $x$-dependent spinor. Alternatively, it can be expressed as a linear combination of the basis spinors  
\begin{equation}
	\label{genSol}
	\psi_{k_1,k_2}(x,y,\tau)= \left(
	\begin{array}{c}
		1 +k_1-\Ti k_2 \\
		-1 -k_1-\Ti k_2 \\
	\end{array}
	\right) e^ {\Ti (k_1 x+k_2 y)-2 k_1 k_2 \tau}\,,\qquad 1=k_1^2+k_2^2\,,\qquad k_1,k_2\in \C\,.
\end{equation}

Let us take the following special choice for the seed matrix,
\begin{equation}
	\Phi=e^{\Ti \frac{\pi}{4}}(\phi,\mathcal{T}\phi)\,,\qquad
	\phi=\psi_{k_1, k_2}+\psi_{k_1,- k_2}-\psi_{-k_1, k_2}-\psi_{-k_1,- k_2}\,,
\end{equation}
where $k_1$ and $k_2$ are parametrized by a real factor $\gamma$ as
\begin{equation}
	k_1=\cosh\gamma\,,\qquad
	k_2=\Ti\sinh\gamma\,,
\end{equation}
and we exclude the case $\gamma=0$, since it only reproduces the constant background case up to a sign. This choice of $\Phi$ gives rise to the matrix $\Sigma=:\Sigma_\gamma$, 
\begin{align}
	\label{Sigmagamma}
	&\Sigma_\gamma(x,y,\tau)=	\frac{\Ti \cosh\gamma  }{d_\gamma(x,y,\tau)} A_\gamma(x,y,\tau)\,,\qquad 
	d_\gamma(x,y,\tau)=\sinh ^2\gamma  \cos x_1+\cos x_0+\cosh ^2\gamma  \cosh x_2
	,\\
	&A_\gamma(x,y,\tau)=\left(
	\begin{array}{cc}
		\sinh \gamma  (\Ti \sinh \gamma  \sin x_1- \cosh \gamma  \sinh x_2) & -\cosh \gamma  \cosh x_2-\cosh (\gamma - \Ti x_0) \\
		-\cosh \gamma  \cosh x_2-\cosh (\gamma +\Ti x_0) & \sinh \gamma  (\Ti \sinh \gamma  \sin x_1+\cosh \gamma  \sinh x_2) \\
	\end{array}
	\right)\,,
\end{align}
where we have introduced the shorthand notation 
\begin{equation}
	\label{notation}
	x_0= 2\tau  \sinh 2 \gamma\,,\qquad x_1= 2x\cosh\gamma\,,\qquad x_2= 2y\sinh\gamma\,.
\end{equation}

\begin{figure}[H]
	\begin{center}
		\includegraphics[scale=0.55]{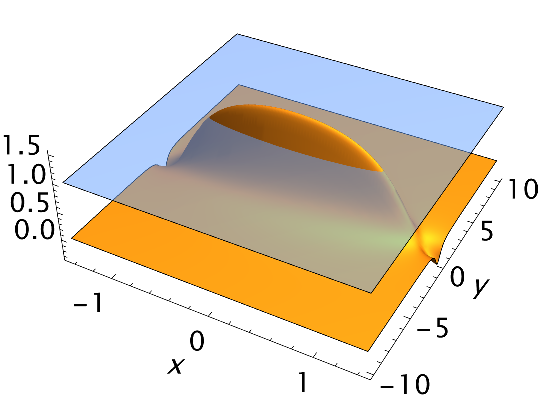}\qquad
		\includegraphics[scale=0.55]{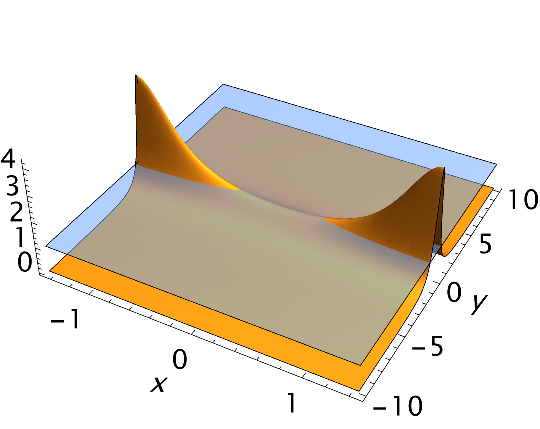}
	\end{center}
	\caption{\small  Plots of $V_\gamma(x,y,\tau)$ and $m_\gamma(x,y,\tau)$ are shown for $\tau=2$ and $\gamma=0.4$. In both panels, the SKT energy level $E=1$ is indicated, illustrating the regions where it intersects and exceeds the potential barrier.}
	\label{figK}
\end{figure}

The corresponding matrix potentials $U_1=U_\gamma$ and $W_1=W_\gamma$ are obtained by introducing (\ref{Sigmagamma}) into equations (\ref{LUW}). For the relation between this example and Dirac systems, we only need the explicit form of $U_\gamma$ (matrix $W_1$ will be relevant for the construction of more general examples in the following sections). The matrix  can be written in the very suggestive form
\begin{equation}
	\label{Ugamma}
U_{\gamma}(x,y,\tau)=-\Ti \sigma_2 (-1 + V_\gamma(x,y,\tau)+\Ti m_\gamma(x,y,\tau)\sigma_3)\,,
\end{equation}
where the real functions $ V_\gamma(x,y,\tau)$ and $ m_\gamma(x,y,\tau)$ are given by 
\begin{align}
	&V_\gamma(x,y,\tau)= \frac{2   \sinh^2\gamma  \left(\cos x_1-\cos x_0 \right)}{d_\gamma(x,y,\tau)}\,, &m_\gamma(x,y,\tau)=-\frac{2\sinh \gamma  \cosh \gamma  \sin  x_0}{d_\gamma(x,y,\tau)}\,.
\end{align}
From here, we can extract the complex amplitude by using the first equation in (\ref{uv}), 
\begin{align}
	&u_1(x,y,\tau)=(-\Ti \sigma_2 U_\gamma)_{2,2}=1-V_\gamma(x,y,\tau)+\Ti\, m_\gamma(x,y,\tau)=: u_\gamma\,.
\end{align}
Being periodic on $x$, the solution lies on the surface of an infinite cylinder parametrized by 
\begin{equation}
	\label{Cylinder}
\Omega=\left\{(x,y)|\quad  x\in \Big[ -\frac{\pi}{2\cosh\gamma},\frac{\pi}{2\cosh\gamma}\Big)\,,
	\quad 	y\in \R \right\}\,,
\end{equation}
while the $\tau$-evolution is also periodic. Thus $u_\gamma(x,y,\tau)$ corresponds to a  solitonic solution periodic in $\tau$, or breather, of the DS II equations \cite{Breather}. Moreover, in the limit  $|y|\rightarrow \pm \infty$, the breather approaches a constant value
\begin{equation}
	\label{limSoli}
	\lim_{y \rightarrow \pm\infty} u_\gamma(x,y,\tau)=1\,.
\end{equation}
Some plots of functions $V_\gamma(x,y,\tau)$ and $m_\gamma(x,y,\tau)$ are shown in Fig. \ref{figK}.

To relate this construction to Dirac-like operators, we must take a step back and focus on the first operator in (\ref{DS-system 0}). If we multiply the equation 
$h_0\psi=0$ by $-\Ti \sigma_2$ from the left then we find
\begin{equation}
	\label{key}
0=	-\Ti \sigma_2h_0\psi=(H_0+1)\psi\qquad\text{with}\qquad
	H_0=-\Ti (\sigma_1\partial_x+\sigma_2\partial_y)\,.
\end{equation}
The operator $H_0$ is a massless free Dirac Hamiltonian in the Euclidean plane (here the Dirac gamma matrices are realized as Pauli matrices). At the same time, the first equation in (\ref{key}) can be reinterpreted as a Schr\"odinger equation $H_0\psi=E\psi$ with energy $E=-1$ (scale transformations can modify this value). Operator $H_0$ enjoys time reversal symmetry,  generated by $\mathcal{T}$,\footnote{The equation $H_\gamma(\tau)\psi=\Ti\frac{\partial}{\partial t}\psi$ is invariant under the action of $\mathcal{T}$, together with the change $t\rightarrow -t$.} as well as reflections in the plane, generated by  
\begin{align}
	\label{Paritx}
	&\mathcal{P}_{x}= \Ti \sigma_2 \pi_x \qquad \pi_x:x\rightarrow-x\,,
	&\mathcal{P}_{y}= \Ti \sigma_1 \pi_y \qquad\pi_y: y\rightarrow-y\,.
\end{align}
The system is also invariant under the action of the Euclidean group, composed of $x$ and $y$ translations, and rotations. 

A similar analysis can be carried out for the transformed operator $h_1=\partial_y-\Ti \sigma_3 \partial_x-U_\gamma=:h_\gamma$. As a result, we obtain
\begin{equation}
	-\Ti \sigma_2 h_\gamma=-\Ti(\sigma_1\partial_{x}+\sigma_2\partial_y)+\Ti \sigma_2U_\gamma=H_\gamma(\tau)-1\,,
\end{equation}
where  
\begin{equation}
	H_\gamma(\tau)=-\Ti(\sigma_1\partial_{x}+\sigma_2\partial_y)+V_\gamma(x,y,\tau)+\Ti \, m_\gamma (x,y,\tau)\sigma_3\,.
\end{equation}
The operator $H_\gamma(\tau)$ is a Dirac Hamiltonian with a nontrivial potential and an imaginary mass term, defined on the cylinder surface. At the same time, $\tau$ plays the role of a free parameter (not to be confused with the Dirac evolution time). Note that the transformation preserves only $\mathcal{T}$ and $\mathcal{P}_{x}\mathcal{P}_{y}$ symmetries, since the individual actions of  $\mathcal{P}_x$ and $\mathcal{P}_y$ change the sign in front of the imaginary mass term. Nevertheless, the system remains $\mathcal{PT}$-symmetric, featuring a non-Hermitian Hamiltonian with real eigenvalues \cite{PT1,PT2}. 

In the particular special case when $\tau=0\Rightarrow m_\gamma=0$, we obtain a massless Hermitian Dirac system given by 
\begin{equation}
	\label{H01}
	H_\gamma(0)=-\Ti(\sigma_1\partial_x+\sigma_2\partial_y) -\frac{4  \sinh ^2\gamma  \sin^2 (x \cosh\gamma)}{\sinh ^2\gamma  \cos (2  x\cosh\gamma)+\cosh ^2\gamma \cosh (2 y\sinh\gamma)+1}\,.
\end{equation}
This system was first constructed in \cite{asym1} by using a Darboux transformation for $(1{+}1)$-dimensional Dirac systems, together with a Wick rotation. In contrast to the case $\tau\not=0$, the model (\ref{H01}) is invariant under $\mathcal{T}$, $\mathcal{P}_x$, and $\mathcal{P}_y$ transformations. On the other hand, it is clear that the Euclidean group is not a symmetry of the model for arbitrary $\tau$. Moreover, due to periodicity, the coordinate $x$ is of the nature of an angle, and the angular momentum is lost. 

To construct the scattering states of $H_\gamma(\tau)$, we introduce $H_0$ and $H_\gamma(\tau)$ in the intertwining relation (\ref{Inter}). As a result, we obtain an asymmetric intertwining relation 
\begin{equation}\label{asym}
	\sigma_2L_\gamma(\tau) \sigma_2 (H_0+1)=(H_\gamma(\tau)-1)L_\gamma(\tau)\,,\qquad L_\gamma(\tau)=\partial_x-\Sigma_\gamma(x,y,\tau)\,.
\end{equation}
In comparison with (\ref{Inter}), the intertwining operator on the left-hand side is dressed now by the $\sigma_2$ matrix, and the Hamiltonians on both sides are shifted by constants with different signs. In consequence, the operator $L_\gamma(\tau)$ only transforms the solutions of $H_0$ with energy $E=-1$ into $H_\gamma(\tau)$ with energy $E=1$, i.e.,

\begin{equation}
	(H_0+1)\psi=0 \qquad \Rightarrow\qquad  (H_1-1)L_\gamma(\tau)\psi=\sigma_2L_\gamma(\tau)\sigma_2\,(H_0+1)\psi=0\,,
\end{equation}
while solutions with $E\not=-1$ can not be obtained using this method,
\begin{equation}
(H_0+1)\psi=\epsilon\psi \qquad \Rightarrow\qquad  (H_1-1)L_\gamma(\tau)\psi=\sigma_2\,L_\gamma(\tau)\,\sigma_2\, (H_0+1)\psi=\epsilon\, \sigma_2L_\gamma(\tau)\sigma_2\psi\,.
\end{equation}
This implies that the system we are constructing is quasi-exactly solvable, albeit not in the traditional sense introduced in the classical literature \cite{Turbiner1}, where the number of exactly solvable states is associated with a finite-dimensional representation of a Lie algebra. In fact, as we will show below, the space of solutions that can be generated at a fixed energy level is infinite-dimensional.

 In particular, the scattering states of $H_\gamma(\tau)$ with $E=1$ are given by
\begin{equation}
	\label{SKTstates0}
	\chi_\gamma (x,y|\tau,\theta)= L_\gamma(\tau)\,\psi(x,y|\theta)\qquad\Rightarrow\qquad  
	H_\gamma(\tau)\,\chi_\gamma (x,y|\tau,\theta)=\,\chi_\gamma (x,y|\tau,\theta)\,,
\end{equation}
where 
\begin{align}
	\label{FreeSpinors}
	&	\psi(x,y|\theta)= e^{\Ti  (x\cos \theta +y\sin\theta )}\left(
	\begin{array}{c}
		1 \\
		-e^{\Ti \theta } \\
	\end{array}
	\right)\,,
\end{align}
is a free-particle plane-wave state with energy $E=-1$, and  $\theta\in(-\pi,\pi)$ denotes  the incident angle. It is straightforward to show that time reversal transformations and reflections acting on the state (\ref{SKTstates0}) only produce a shift $\theta\rightarrow \theta +\pi$, together with some irrelevant proportionality constant.

Since the soliton $u_\gamma(x,y,\tau)$ asymptotically approaches a constant that coincides with the energy of the scattering state, see (\ref{limSoli}), the Dirac system $H_\gamma(\tau)$ is asymptotically free, 
\begin{equation}
	\lim_{y\rightarrow\pm \infty}V_\gamma(x,y,\tau)=0\,,\qquad
	\lim_{y\rightarrow\pm \infty}m_\gamma(x,y,\tau)=0\,.
\end{equation}
Consequently, the asymptotic behaviour of $\chi_\gamma (x,y|\tau,\theta)$ is decomposed in terms of incident, reflected and transmitted free-fermion states with  $E=1$, 
\begin{align}
	\label{as-inf}
	&\lim_{y\rightarrow  -\infty}\chi_\gamma (x,y|\tau,\theta) =c_-\,e^{\Ti (x\cos \theta +y\sin\theta )}\left(
	\begin{array}{c}
		1 \\
		e^{\Ti \theta } \\
	\end{array}
	\right) +c_0\,e^{\Ti (x\cos \theta -y\sin\theta )}\left(
	\begin{array}{c}
		e^{\Ti \theta }  \\
		1\\
	\end{array}
	\right)\,,\\
	\label{as+inf}
	&	\lim_{y\rightarrow  \infty}\chi_\gamma (x,y|\tau,\theta) =c_+\,e^{\Ti (x\cos \theta +y\sin\theta )}\left(
	\begin{array}{c}
		1 \\
		e^{\Ti \theta } \\
	\end{array}
	\right)\,.
\end{align} 
In terms of the constants $A$, $B$ and $C$ we can introduce the reflection and the transmission amplitude as
\begin{equation}
	r=\frac{c_0}{c_-}\,,\qquad 	t=\frac{c_+}{c_-}\,,\qquad 
	|r|^2+|t|^2=1\,.
\end{equation}

\begin{figure}[H]
	\begin{center}
		\includegraphics[scale=0.49]{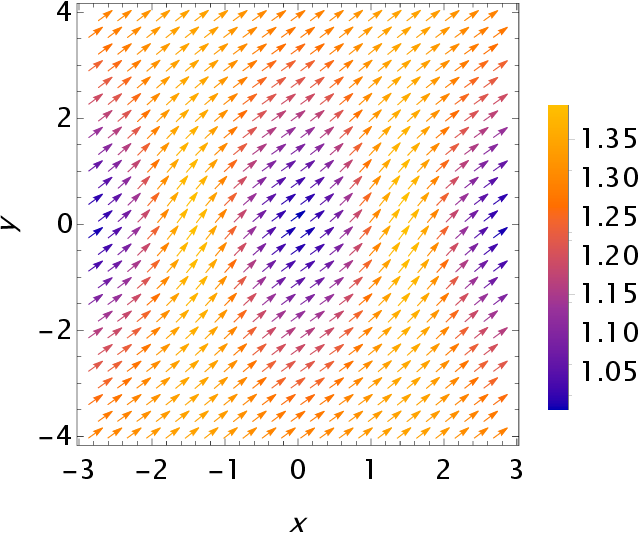}
		\includegraphics[scale=0.49]{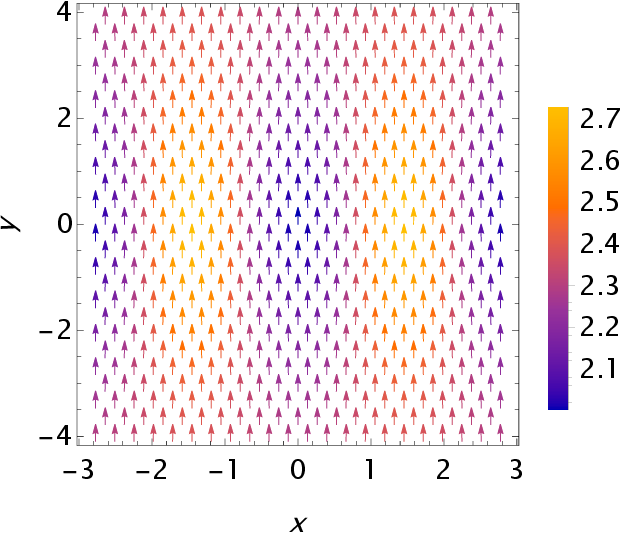}
		\includegraphics[scale=0.49]{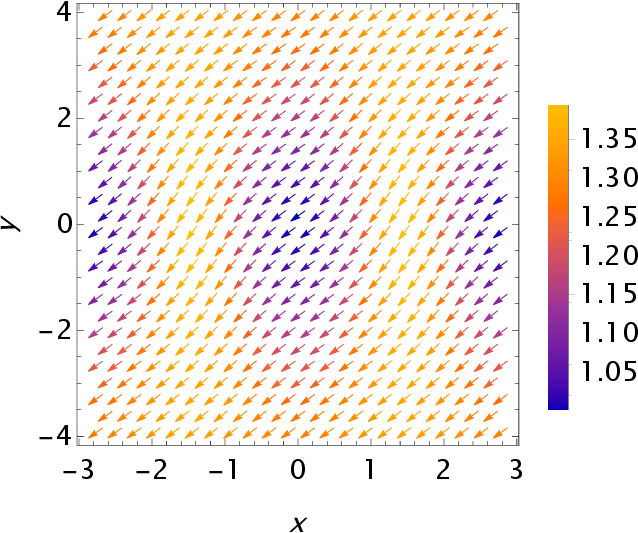}
	\end{center}
	\caption{\small   Probability density current of the SKT states plotted over two spatial periods i.e., $x\in(-\sech\gamma,\sech\gamma)$, in order to better illustrate the effect of periodic boundary conditions. The parameters are set to  $\tau=0$ (Hermitian case) and $\gamma=0.4$, while the incident angles are (from left to right) given by $\theta=\{\frac{\pi}{4},\frac{\pi}{2},\frac{5\pi}{4}\}$. The maximum (minimum) probability corresponds to the brightest (darker) arrows in each figure. }
	\label{SKTcurrent1}
\end{figure}

\begin{figure}[H]
	\begin{center}
		\includegraphics[scale=0.5]{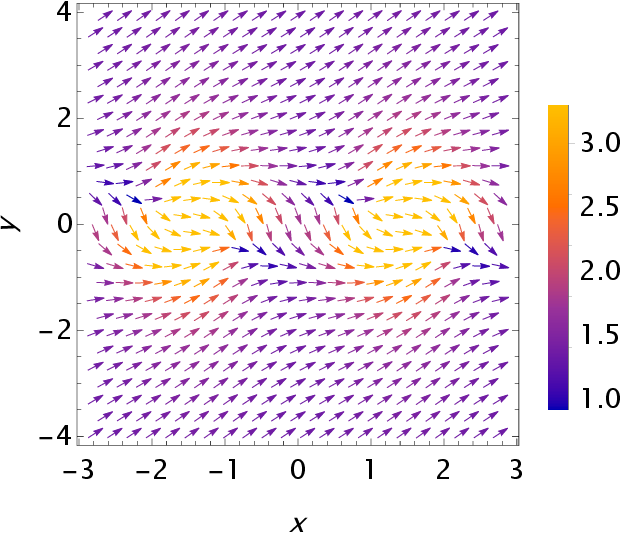}
		\includegraphics[scale=0.5]{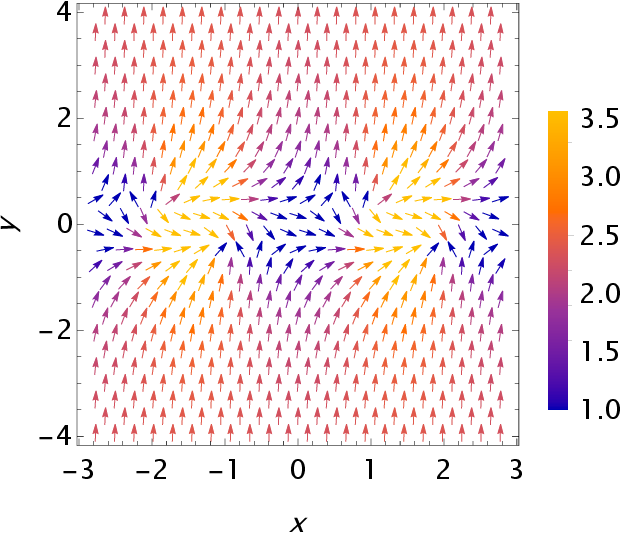}
		\includegraphics[scale=0.5]{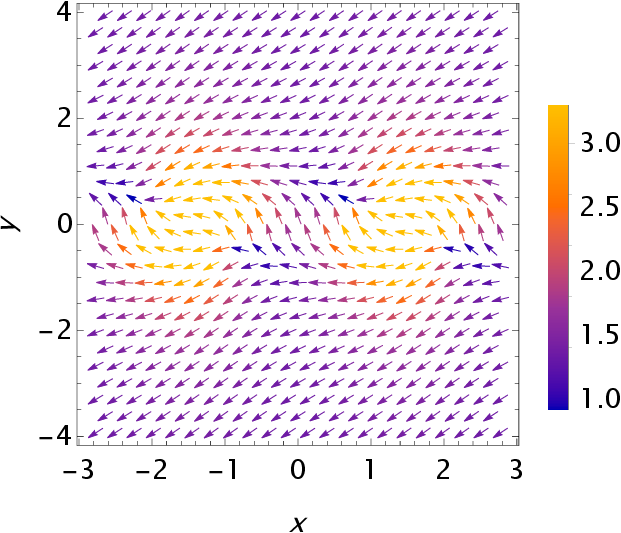}
	\end{center}
	\caption{\small  Probability density current of the SKT states for $x\in(-\sech\gamma,\sech\gamma)$.  The parameters are set to $\tau=2$ (non-Hermitian) and $\gamma=0.4$. The incident angles are (from left to right) $\theta=\{\frac{\pi}{4},\frac{\pi}{2},\frac{5\pi}{4}\}$. The maximum (minimum) probability corresponds to the brightest (darker) arrows in each figure.}
	\label{SKTcurrent2}
\end{figure}

To investigate the scattering process, we compute the form of the intertwining operator in the limits $y\rightarrow\pm \infty$, since $\lim_{y\rightarrow \pm \infty}\chi_\gamma (x,y|\tau,\theta)=L_\pm\psi(x,y|\theta)$, 
\begin{equation}
	\label{asymp}
	L_\pm:=\lim_{y\rightarrow \pm \infty}L_\gamma(\tau)=\partial_x -\Sigma_\pm\qquad\text{with}\qquad  \Sigma_\pm=\Ti  ( \mp  \sinh\gamma \sigma _3-\sigma _1  )\,.
\end{equation}
This operator does not change the plane-wave behavior of the states, it does not depend on $\tau$, and its action on (\ref{FreeSpinors}) reproduces the relations (\ref{as-inf}) and (\ref{as+inf}) with 
\begin{equation}
	\label{transmisocoe}
	c_\pm=\sin\theta\pm\Ti \sinh\gamma\,,\qquad c_0=0\,,\qquad r=0\,,\qquad 
	t(\theta)=\frac{\sin\theta+\Ti \sinh\gamma}{\sin\theta-\Ti \sinh\gamma}\,.
\end{equation}
 This result indicates that the wave function undergoes a phase shift as it passes through the interaction region, without being reflected. The transmission coefficient is unity, $|t(\theta)|^2=1$. This feature is the celebrated SKT effect, corresponding to total transmission at any incidence angle of the wave \cite{SKT1,SKT2,asym1}. For this reason, we refer to $\chi_\gamma (x,y|\tau,\theta) $ 
 as the \textit{SKT states} hereafter. The effect can be observed in the plots of the probability density current for these states, 
\begin{equation} \vec{j} = (j_1,j_2)  \quad \text{ with} \quad  j_k(x,y|\tau,\theta)=\chi_\gamma (x,y|\tau,\theta)^\dagger \sigma_k \chi_\gamma (x,y|\tau,\theta) \quad \text{for} \quad k=1,2
\end{equation}  
in both the Hermitian $(\tau=0)$ and $\mathcal{PT}$-symmetric ($\tau\not=0$) cases, see Fig. \ref{SKTcurrent1} and Fig. \ref{SKTcurrent2}.
In the Hermitian scenario, probability is conserved, and the particle undergoes a phase shift as it traverses the potential. In contrast, in the $\mathcal{PT}$-symmetric case, the continuity equation for probability incorporates a source or sink term proportional to the imaginary component of the mass in the Hamiltonian,
\begin{equation}
\vec\nabla \cdot \vec{j} =2 m_{\gamma}(x,t,\tau)\chi_\gamma (x,y|\tau,\theta)^\dagger \sigma_3\chi_\gamma (x,y|\tau,\theta).
\end{equation}
This extra term can be interpreted as a local counterflow, leading to destructive interference that suppresses potential backscattering. Consequently, the vertical component $j_2$ of the current may vanish locally, producing stagnation points, as the incident flux is redirected into transverse modes and locally balanced by the non-Hermitian gain-and-loss mechanism. In the asymptotic regime, this combined mechanism restores the probability flux in the forward direction, yielding unit transmission accompanied by a characteristic phase shift $\delta(\theta)=\arg(t(\theta))$.

It is well known in quantum mechanics that poles in the transmission amplitude indicate the existence of bound states. This occurs when the incident angle takes the imaginary value $\theta = \Ti \gamma$, corresponding to a single bound state in the continuum \cite{BIC}. By applying the discrete symmetries, we produce more bound states. In total,  
\begin{align}
	\label{bound1}
	\zeta_\gamma^{(1)}(x,y|\tau)=\chi_\gamma (x,y|\tau,\Ti\gamma)&=L_\gamma(\tau)\psi(x,y|\Ti\gamma)\,,\\
	\zeta_\gamma^{(2)}(x,y|\tau)=\mathcal{T}\chi_\gamma (x,y|\tau,\Ti\gamma)&=L_\gamma(\tau)\mathcal{T}\psi(x,y|\Ti\gamma)\,,\\
	\zeta_\gamma^{(3)}(x,y|\tau)=
	\mathcal{P}_x\mathcal{P}_y\chi_\gamma (x,y|\tau,\Ti\gamma)&=-L_\gamma(\tau)\mathcal{P}_x\mathcal{P}_y\psi(x,y|\Ti\gamma)\,,\\
	\zeta_\gamma^{(4)}(x,y|\tau)=
	\mathcal{T}\mathcal{P}_x\mathcal{P}_y\chi_\gamma (x,y|\tau,\Ti\gamma)&=-L_\gamma(\tau)\mathcal{T}\mathcal{P}_x\mathcal{P}_y\psi(x,y|\Ti\gamma)\,.
	\label{bound4}
\end{align}
Since these four states are related by discrete symmetries of the Hamiltonian, they have the same probability density given by
\begin{equation}
	\label{RhoPT}
	\rho_\gamma(x,y,\tau)=\zeta_\gamma^{(a)} (x,y|\tau)^\dagger \zeta_\gamma^{(a)} (x,y|\tau)=\frac{N_\gamma^2}{\sinh^2\gamma\cos  x_1 +\cos  x_0 +\cosh ^2\gamma  \cosh x_2}\,,\qquad a=1,2,3,4\,,
\end{equation}
where $N_\gamma^2$ is the normalization constant defined by the equation $	\int_\Omega dxdy 	\rho_\gamma(x,y,\tau)=1$. 

All these calculations were performed under the assumption that $\tau$ is real, recalling its previous role as a soliton evolution parameter. However, within the context of the Dirac systems, we have the freedom to treat this parameter as an imaginary number, i.e.,  $\tau = \Ti z$. If we do this, our Hamiltonian becomes Hermitian for any real value of $z$,
\begin{align}
	&	H_\gamma (\Ti z)= -\Ti(\sigma_1\partial_{x}+\sigma_2\partial_y)+\widetilde{V}_\gamma(x,y,z)+ \, \widetilde{m}_\gamma (x,y,z)\sigma_3=:\tilde{H}_\gamma(z)\,,\\
	&\widetilde{V}_\gamma(x,y,z)= \frac{2  \sinh^2\gamma  \left(\cosh x_3 +\cos x_1\right)}{\sinh^2\gamma\cos  x_1 +\cosh  x_3 +\cosh ^2\gamma  \cosh x_2}\,,
	\\&\widetilde{m}_\gamma(x,y,z)=\frac{\sinh 2 \gamma   \sinh  x_3}{\sinh^2\gamma\cos  x_1 +\cosh  x_3 +\cosh ^2\gamma  \cosh x_2}\,.
\end{align}
where we have used the notation 
\begin{equation}
\label{x3}
x_3=2\,z  \sinh 2 \gamma.
\end{equation}
The price paid for considering such a system is the loss of time-reversal symmetry. Indeed, the action of this operator now induces a change of sign in $z$,
\begin{equation}
\label{Tinter}
	\mathcal{T}\tilde{H}_\gamma( z)\mathcal{T}^\dagger=\widetilde{H}_\gamma(- z)
	\,,\qquad
	\mathcal{T}\widetilde{L}_\gamma(z)\mathcal{T}^\dagger=\widetilde{L}_\gamma(- z)\,,\qquad \widetilde{L}_\gamma(z):=L_\gamma(\Ti z)\,.
\end{equation} 

Despite this issue,  the scattering properties do not depend on $\tau$, see (\ref{asymp}). Thus, the new Hermitian systems also have SKT states, constructed by simply replacing $\tau$ with $\Ti z$ in  (\ref{SKTstates0}). One can also check by direct computation that the probability density current for SKT state is conserved for any real $z$. However, the four bound states constructed as in (\ref{bound1})-(\ref{bound4}) but with $\tau\rightarrow \Ti z$,
\begin{align}
	\label{boundHer1}
	&\zeta_\gamma^{(1)}(x,y|\Ti z)=\widetilde{L}_\gamma(z)\psi(x,y|\Ti\gamma)\,,\\
	&	\zeta_\gamma^{(2)}(x,y|\Ti z)=\widetilde{L}_\gamma(z)\mathcal{T}\psi(x,y|\Ti\gamma)=
	\mathcal{T}\widetilde{L}_\gamma(z)\psi(x,y|\Ti\gamma)=\mathcal{T}\zeta_\gamma^{(1)}(x,y|-\Ti z)
	\,,\\
	&	\zeta_\gamma^{(3)}(x,y|\Ti z)=-\widetilde{L}_\gamma(z)\mathcal{P}_x\mathcal{P}_y\psi(x,y|\Ti\gamma)=\mathcal{P}_x\mathcal{P}_y\widetilde{L}_\gamma(z)\psi(x,y|\Ti\gamma)=\mathcal{P}_x\mathcal{P}_y\zeta_\gamma^{(1)}(x,y|\Ti z)\,,\\
	&	\zeta_\gamma^{(4)}(x,y|\Ti z)=-\widetilde{L}_\gamma(z)\mathcal{P}_x\mathcal{P}_y \mathcal{T}\psi(x,y|\Ti\gamma)=\mathcal{P}_x\mathcal{P}_y \mathcal{T}\widetilde{L}_\gamma(z)\psi(x,y|\Ti\gamma)=\mathcal{P}_x\mathcal{P}_y \mathcal{T}\zeta_\gamma^{(1)}(x,y|-\Ti z)\,,\label{boundHer4}
\end{align}
have pairwise equal probability density, 
\begin{align}
	\label{RhoHer}
	&\rho_\gamma^{(+)}(x,y, z)=(\zeta_\gamma^{(1)}(x,y| \Ti z))^\dagger (\zeta_\gamma^{(1)}(x,y|\Ti z))=(\zeta_\gamma^{(3)}(x,y| \Ti z))^\dagger (\zeta_\gamma^{(3)}(x,y|\Ti z))\,,\\
	&\rho_\gamma^{(-)}(x,y, z)=(\zeta_\gamma^{(2)}(x,y| \Ti z))^\dagger (\zeta_\gamma^{(2)}(x,y|\Ti z))=(\zeta_\gamma^{(4)}(x,y| \Ti z))^\dagger (\zeta_\gamma^{(4)}(x,y|\Ti z))\,,
	\\& 
	\rho_\gamma^{(+)}(x,y, z)=\rho_\gamma^{(-)}(x,y, -z)\,.
\end{align}

\begin{figure}[H]
	\begin{center}
		\includegraphics[scale=0.5]{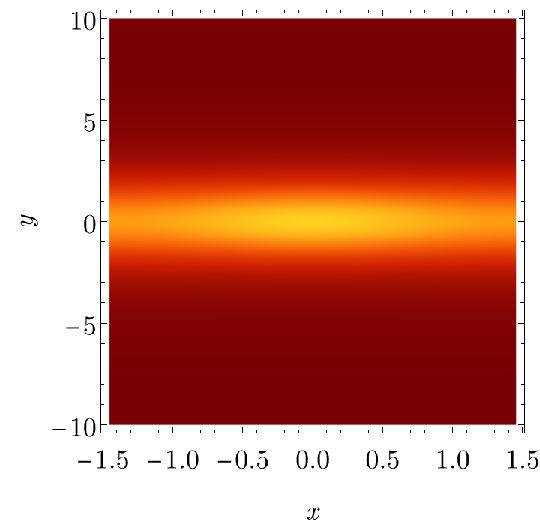}
	    \includegraphics[scale=0.5]{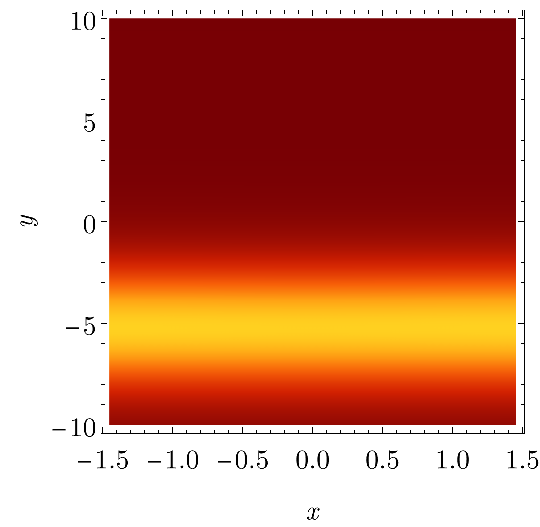}
		\includegraphics[scale=0.5]{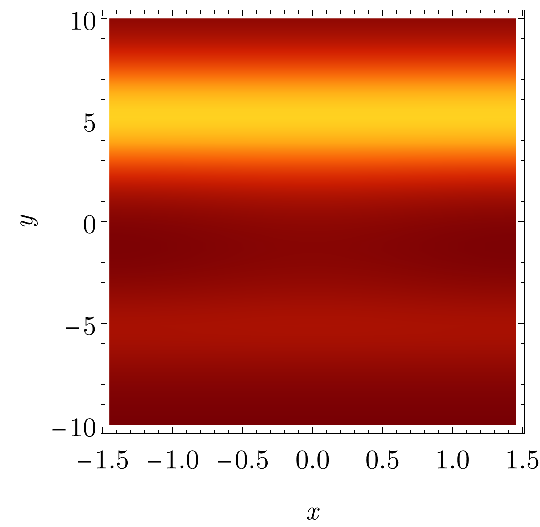}
	\end{center}
	\caption{\small Differences in the probability density for bound states in the $\mathcal{PT}$-symmetric and the Hermitian cases. In the first panel, the density plot for (\ref{RhoPT}) is shown. Subsequently, we display the probability density (\ref{RhoHer}) with plus and minus signs (in that order) for the Hermitian case. Maximal (minimal) brightness corresponds to maximal (minimal) probability of finding the confined particle described by the bound state. The parameters are  $\gamma=0.4$, $\tau=2$ and $z=2.5$. }
	\label{ProbPTbound}
\end{figure}

In contrast to the $\mathcal{PT}$-symmetric case, particles described by bound states in the Hermitian case show a preference to be found in the upper or lower half-plane bounded by the $y$ axis, depending on the sign in $\rho_\gamma^{(\pm)}(x,y, z)$. This comparison is shown in Fig. \ref{ProbPTbound}.

Up to this point we have presented a specific example among a large set of solutions, as there are infinitely many ways to construct seed matrices $\Phi$, and therefore infinitely many DS II solutions.  In the following section, we will explore the necessary conditions for building the most general DS II solution that enables the construction of a Dirac operator exhibiting the SKT effect.

\section{The SKT/DS II relation: A proper choice of seed}
\label{Sec4}
In order to construct the most general system exhibiting the SKT effect through its connection with DS II solutions, one is naturally led to the inverse problem \cite{insk}, namely, the reconstruction of a DS II solution starting from the SKT boundary conditions. In the present work, however, we address a slightly different question: what is the most general Dirac system displaying the SKT effect that can be generated via a first-order Darboux transformation, using as a seed the constant DS II solution (\ref{DS-system 0})?

Let us first determine which boundary conditions guarantee the existence of the SKT effect. By looking at (\ref{asymp}) and (\ref{transmisocoe}), we note that the key structure in this regard is the $\Sigma$ matrix, which must be such that the spinors do not have a reflecting part. Inspired by that, we look for the case where $\Sigma$ tends to a constant matrix $\Sigma_\pm$, invariant under $\mathcal{T}$, when looking beyond the origin, i.e.,
\begin{align}
&\label{BCS}
	\lim_{y\rightarrow \pm \infty}\Sigma =:	\left(\begin{array}{cc}
	a_\pm e^{\Ti \alpha_\pm } & b_\pm e^{\Ti \beta\pm }\\
	 -b_\pm e^{-\Ti \beta_\pm} & a_\pm  e^{-\Ti \alpha_\pm} 
	\end{array}\right)=\Sigma_\pm \,,\qquad a_\pm,b_\pm,\alpha_\pm,\beta_\pm\in \R\,,\\
& 	\lim_{y\rightarrow \pm \infty}\partial_x\Sigma=\lim_{y\rightarrow \pm \infty}\partial_y\Sigma=0\,.
\end{align}
The constant parameters in (\ref{BCS}) are restricted by the requirement 
\begin{equation}
\label{KleinBC}
\lim_{y\rightarrow \pm \infty}\,L\psi_\theta=(\partial_x-\Sigma_\pm)\psi_\theta= c_\pm e^{\Ti (x\cos \theta +y\sin\theta )}\left(
\begin{array}{c}
 1 \\
 e^{\Ti \theta } \\
\end{array}
\right)\,,\qquad t=\frac{c_+}{c_-}
\end{equation}
where $c_-=\overline{c_+}$ are self-conjugate complex constants. To fix these free parameters, we develop a differential equation for the matrix $\Sigma$. To this end, we recall the matrix potential $W_1$, which can be expressed in terms of derivatives of $\Sigma$ (see (\ref{LUW})) or in terms of the fields (see (\ref{W0})),
\begin{equation}
W_1=\tfrac{1}{2}(w_1 +\Ti \sigma_3Q_1)+\Ti (\partial_y+\Ti \sigma_3\partial_x)U_1\sigma_3= W_0+2(\partial_y+\Ti\sigma_3\partial_x)\Sigma\,.
\end{equation}
By reorganizing the terms, we derive the differential equation 
\begin{equation}
(\partial_y+\Ti \sigma_3\partial_x)(2\Sigma-\Ti U_1\sigma_3)=\tfrac{1}{2}(w_1+\Ti \sigma_3\partial_x Q_1)-W_0\,.
\end{equation}
Let us now apply $(\partial_y-\Ti \sigma_3\partial_x)$ from the left to get
\begin{equation}
\begin{aligned}
(\partial_x^2+\partial_y^2)(2\Sigma-\Ti U_1\sigma_3)
&=\Ti (\partial_y+\Ti \sigma_3\partial_x) U_1^2\sigma _3
-(\partial_y-\Ti \sigma_3\partial_x)W_0\,,
\end{aligned}
\end{equation}
 When $W_0=0$, corresponding to the case (\ref{consstantback}), the last equation reduces to a total derivative so that it can be integrated once.  After some algebraic manipulations, we get a kind of Riccati equation for $\Sigma$,
 \begin{equation}
 	\label{Matrix Ricatti}
 	(\partial_y-\Ti \sigma_3\partial_x)(U_0+\Ti \{\sigma_3,\Sigma\})+(U_0+\Ti[\sigma_3,\Sigma])^2=C(\tau)\,,
 \end{equation} 
where we have used $U_1=U_0+\Ti[\sigma_3,\Sigma]$. Remarkably, the left-hand side of this equation is covariant with respect to $\tau$ evolution generated by the operator $e^{-2\tau \partial_x\partial_y}$, while the right-hand side is invariant under such a transformation. Therefore, $ C$ must be a fixed constant, and the $\tau$ dependence is not relevant for the study of the asymptotic behavior. To fix the constant value we use the definition (\ref{Sigma}) of $\Sigma$ and obtain
 \begin{equation}
 C=U_0^2=-\sigma_0\,.
 \end{equation}
Considering (\ref{Matrix Ricatti}) in the limits $y\rightarrow\pm \infty$ together with the boundary condition (\ref{BCS}) we see that  
\begin{equation}
	\label{BC2}
\lim_{y\rightarrow \infty}U_1^2=(-\Ti  \sigma_2+\Ti[\sigma_3,\Sigma_\pm])^2= -\sigma_0\,.
\end{equation}  
Unlike the usual picture of the inverse scattering problem, where solitons are assumed to vanish far from the origin of the coordinates, the relation (\ref{BC2}) means that the DS-II solitons associated with reflectionless scattering are characterized by an asymptotic constant, i.e., $\lim_{y\rightarrow \pm \infty}|u_1|=1$. The relation (\ref{BC2}) is also useful to compute the explicit form of $b_\pm$, which gives us 
 \begin{equation}
 b_{\pm}=- \sin\beta_\pm\,.
\end{equation}  
By using (\ref{KleinBC}) we find that the other constants must be chosen such that  
\begin{equation}a_\pm =\sinh\gamma\,,\qquad \alpha_\pm=\mp \frac{\pi}{2}\,,\qquad \beta_\pm=\frac{\pi}{2}\,,\qquad\Rightarrow\qquad
	\Sigma_\pm= \Ti (\mp\sinh\gamma \sigma_3-\sigma_1)\,.
\end{equation}
This reveals that the condition for Klein tunneling has a  unique form for breather solutions of DS II, and the transmission amplitude is given by (\ref{transmisocoe}). 

From the asymptotic behavior for $\Sigma_\pm$, we can also obtain the boundary conditions for the seed solutions. This can be done by using the equation $L\Phi=0$ in the limit $y\rightarrow \pm \infty$, i.e., we must integrate the equation $(\partial_x -\Sigma_\pm)\Phi_\pm=0$. This gives us
\begin{equation}
\begin{aligned}
\Phi_\pm(x)=\exp(\Sigma_\pm x)C_\pm
= \left(
\begin{array}{cc}
 \cosh (\gamma -\Ti   x \cosh \gamma ) & \mp \Ti \sin ( x \cosh \gamma ) \\
 \mp \Ti \sin ( x \cosh \gamma ) &  \cosh (\gamma +\Ti  x \cosh \gamma ) \\
\end{array}
\right)C_\pm \,,
\end{aligned}
\end{equation}
where $C_\pm$ is, at this point, some arbitrary complex matrix. Indeed, a remarkable feature of this result is that the full dependence of the asymptotic solutions on $x$, $y$, and $\tau$ is obtained by first applying the operator $e^{\Sigma_\pm x}$, followed by the sequential action of the operators in~(\ref{genSol}) on the arbitrary constant matrix $C_\pm$, as follows
\begin{equation}
C_\pm \xrightarrow[e^{\Sigma_\pm x}] {}\Phi_\pm(x)  \xrightarrow[e^{\Ti(\sigma_3\partial_x-\sigma_2)y}] {} \Phi_\pm(x,y) \xrightarrow[e^{2 \tau \partial_x\partial_y}] {}  \Phi_\pm(x,y,\tau)\,.
\end{equation}
In the second step, the dependence on  $y$,  yields  
\begin{equation}
\Phi_\pm(x,y)=e^{\Ti(\sigma_3\partial_x-\sigma_2)y}\Phi_\pm(x)= e^{\pm \sinh\gamma y}\Phi_\pm(x)\,.
\end{equation}
In the same way, we can reconstruct the $\tau$ evolution,
\begin{equation}
	\Phi_\pm(x,y,\tau)= e^{2 \tau \partial_x\partial_y}\Phi_\pm(x,y)=\Phi_\pm(x\pm 2 \tau\sinh\gamma ,y)\,.
\end{equation}
Finally, the most general seed solution that combines both asymptotic behaviors reads 
\begin{equation}
	\label{seedmatrix}
\Phi(x,y,\tau)=\Phi_+(x+2 \tau \sinh\gamma,y)+ \Phi_-(x-2 \tau \sinh\gamma,y)+ e^{2 \tau \partial_x\partial_y}\mathcal{F}(x,y)\,,
\end{equation}
where $\mathcal{F}(x,y)$ is some localized wave packet that vanishes at large distances,
\begin{equation}
\mathcal{F}(x,y)=e^{-\Ti (\sigma_3\partial_y+\sigma_1)x}\mathcal{F}(y)\,,\qquad \lim_{y \rightarrow \pm\infty}\mathcal{F}(y)=0\,.
\end{equation}
When constructing (\ref{seedmatrix}), the condition (\ref{condition}) forces us to restrict the matrices to be $\mathcal{T}$ invariant, i.e., 
\begin{equation}
	\label{Condition2}
\mathcal{T}C_\pm \mathcal{T}^{-1}=C_\pm\,,\qquad
\mathcal{T}\mathcal{F}(x,y) \mathcal{T}^{-1}=\mathcal{F}(x,y)\,.
\end{equation}
With the seed matrix established,  the matrices $\Sigma$, $U_1$, and $W_1$, as well as the solutions to the DS II system, can be derived straightforwardly from the relationships outlined in Section \ref{Sec1}. By construction, the asymptotic behavior of the solution is set to a  constant, as described in (\ref{BC2}). Therefore, we can always express the complex amplitude as 
\begin{align}
\label{GenericSoliton}
&u_1(x,y,\tau)=1 - V(x,y,\tau)+\Ti\, m(x,y,\tau)\,,\\
\label{GenericVandM}
&V(x,y,\tau)=1- \text{Re}(u_1(x,y,\tau))\,,\quad m=\text{Im}(u_1(x,y,\tau))\,,
\end{align}
so that we identify the real functions $V(x,y,\tau)$ and $m(x,y,\tau)$ with the electrostatic potential and the position-dependent mass term, respectively,  of a generic  $\mathcal{PT}$-symmetric Dirac Hamiltonian, 
\begin{align}
&
	H_1=-\Ti(\sigma_1\partial_{x}+\sigma_2\partial_y)+V(x,y,\tau)+\Ti \, m (x,y,\tau)\sigma_3\,.
\end{align}
This Hamiltonian reveals itself to be Hermitian in the case $\tau=0$ only if
\begin{equation}
	\label{initialcond}
	m(x,y,0)=0\,.
\end{equation}
The construction of the SKT and the bound states follows along the lines described in Section \ref{SecExp}. While the asymptotic properties of the states are largely fixed, the behavior of the particles in the interaction region can be varied by tuning the free parameters in the seed solution. In the following section, we construct some explicit examples. 

\section{Parametric family of solutions}
\label{SubSecFam}
In this section, a parametric family of DS~II solutions is constructed using (\ref{seedmatrix}), while neglecting the contribution of an additional wave packet by setting $\mathcal{F}(x,y)=0$. Using (\ref{GenericVandM}), we then derive Dirac Hamiltonians exhibiting the SKT effect. Subsection \ref{SubSecT0} is devoted to the construction at $\tau=0$. In Subsection \ref{SubSecTnot0}, real $\tau$ evolution is included, leading to complex but $\mathcal{PT}$-symmetric systems, as discussed in Section \ref{SecExp}. Finally, Subsection \ref{SubSecTim} analyzes imaginary $\tau$ evolution, which results in Hermitian Dirac models that do not possess time-reversal symmetry.

\subsection{Two-parametric Hermitian family at $\tau=0$}
\label{SubSecT0}

To begin, we examine in detail the structure of the seed matrix itself, whose explicit form is given by
\begin{equation}
\Phi(x,y,0)=   e^{  \sinh\gamma y}e^{\Sigma_+ x}C_++e^{-  \sinh\gamma y}e^{\Sigma_- x}C_-\,.
\end{equation}
At first glance, we have eight free parameters corresponding to the elements of the constant matrices $C_\pm$, but, by means of  $y$-translations and equation (\ref{SimB}), one can reduce the seed matrix to a more basic form. In fact, it is easy to see that 
\begin{align}
\label{ytrans}
&\Phi(x,y+y_0,0)=\left[e^{\Sigma_+ x} e^{\sinh\gamma y}C^\prime_++e^{-\sinh\gamma y}e^{\Sigma_- x}\right]C_-e^{-\sinh\gamma y_0}\,,
\\
&C^\prime_+(0)= e^{ 2\sinh\gamma y_0}C_+C_-^{-1}\,.
\end{align}
By virtue of equation (\ref{SimB}), we note that both the seed matrix $\Phi$ and the term within the square brackets in (\ref{ytrans}) yield the same $\Sigma$ matrix, and thus the same solution. This result indicates the presence of redundant parameters. Therefore, $C_-$ can be fixed as the identity from the beginning without loss of generality. The constant $y_0$ is chosen such that $C^\prime_+\in SU(2)$, in order to satisfy condition (\ref{condition}),
\begin{align}
	\label{seedfinal}
	&	\Phi_\vs(x,y,0)=  e^{ \sinh\gamma y}e^{\Sigma_+x}\mathcal{S}+e^{- \sinh\gamma y}e^{\Sigma_-x}\,,\qquad 
	\mathcal{S}=\left(
	\begin{array}{cc}
		s_1+\Ti s_4 & s_2+\Ti s_3 \\
		-s_2+\Ti s_3 & s_1-\Ti s_4 \\
	\end{array}
	\right)\,,
	\\
	& \det(\mathcal{S})=\sum_{i=1}^4s_i^2=1\,, \qquad s_i\in \R\,. 
\end{align} 

\begin{figure}[H]
\begin{center}
	\includegraphics[scale=0.3]{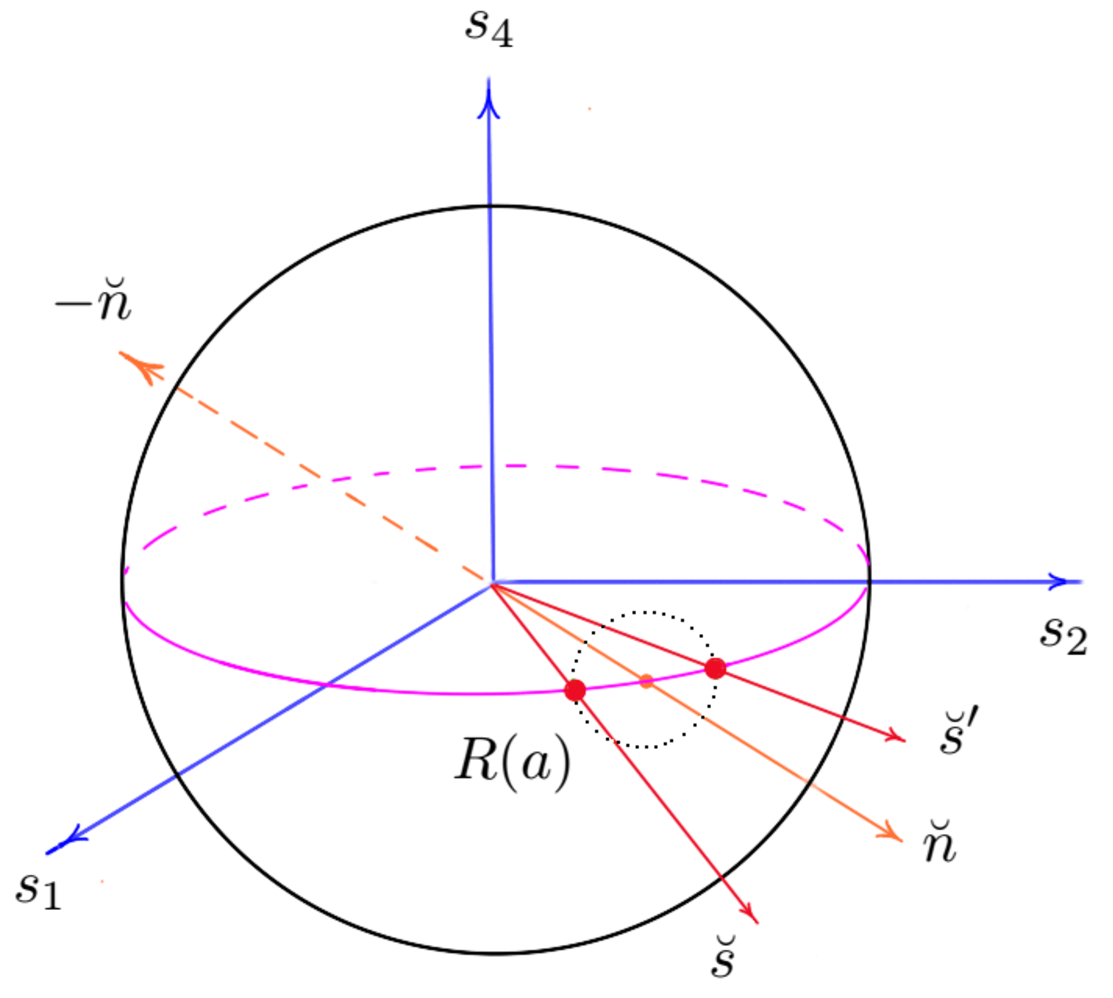}
\end{center}
\caption{\small The parameter space projected to the $S^2(s_3)$ sphere with $s_3=0$. The vector $\breve{n} $, as given in equation (\ref{xtranslation}), defines the rotation axis of $R(a)$. The orbits of $R(a)$, shown as a black dotted circle, represent the equivalence classes of DS II solutions that are related by translations. Bold (red) points stand for two equivalent vectors under the action of $R(a)$, corresponding to half the period of the rotation. The class representatives are chosen in the equatorial semicircle defined between points $\breve{n} $ and -$\breve{n} $. }
\label{FigDiag}
\end{figure}

This choice turns the parameter space into an $S^3$ sphere, with elements represented by the four-dimensional vector $\vs = (s_1, s_2, s_3, s_4)$. Similar to what happened for the $y$-translations, $x$-translations also help in reducing the number of parameters.  Indeed, such a transformation acting on the seed matrix $\Phi_\vs(x,y,0)$ produces
\begin{equation}
	\Phi_\vs(x+a,y,0)=\left[ e^{  \sinh\gamma y}e^{\Sigma_+x}\mathcal{S}_a+e^{- \sinh\gamma y}e^{\Sigma_-x}\right]\,e^{a\Sigma_-}=\Phi_{\vs(a)}(x,y,0)\,e^{a\Sigma_-}\,,\qquad \mathcal{S}_a=e^{a \Sigma_+} \mathcal{S} \,e^{-a\Sigma_-}\,,
\end{equation}
From here we read that the $x$-translations manifest themselves as rotations in the parameter space as follows,
\begin{align}
	\label{xtranslation}
	s_i\rightarrow s_i(a)= R_{ij}(a)s_j\,,\quad R(a)=e^{-2 a\cosh \gamma X} \,,\quad 
	X=
	\left(\begin{array}{cccc}
		0 &0&0& \tanh\gamma \\
		0 &0&0& -\sech\gamma\\
		0 &0&0& 0\\
		-\tanh\gamma &\sech\gamma&0&   0
	\end{array}\right)\,.
\end{align}	
We see that these rotations do not touch $s_3$. To visualize their action, it is therefore sufficient to consider the two-sphere $S^2(s_3)\subset S^3$ for the fixed value $s_3$ given by 
\begin{equation}
	\breve{s}=(s_1,s_2,s_4)\qquad \text{with}\qquad s_1^2+s_2^2+s_4^2=1-s_3^2\,.
\end{equation}
Here the action of $R(a)$ is a rotation about an axis given by the unit vector 
\begin{equation}
	\breve{n}=(\sech\gamma,\tanh\gamma,0)\,,
\end{equation}
and so, the orbits are circle centered around $(1-s_3^2)\breve{n}$. These orbits define equivalence classes of soliton solutions under $x$-translations. The special points $\breve{s}_\pm=\pm (1-s_3)\breve{n}$ in the $s_4=0$ equatorial plane constitute singular orbits, since they are stable under $x$-translations. Varying $\gamma\in \R$ the unit vector $\breve{n}$ rotates in the equatorial plane along the semicircle between $(0,-1,0)$ and $(0,1,0)$, see Fig. \ref{FigDiag}. 

Clearly, for any given value of $\gamma$, one can rotate an arbitrary point $\vec{s}$ along a translational orbit to an equatorial point by choosing \footnote{This equation has two solutions for $a$. The two equatorial points are related by $R\left(\frac{\pi}{2\cosh\gamma}\right)$.} 
\begin{equation}
	\tan(2 a \cosh\gamma)=\frac{s_4 \cosh\gamma}{s_2-s_1\sinh\gamma}\,,
\end{equation}  
and hence move to an orbit representative with $s_4=0$. From now on, we thus set $s_4=0$ and reduce the parameter space to 
\begin{equation}
		\label{ParameterSpace}
	\vs\rightarrow \vec{s}=(s_1,s_2,s_3)\in S^2(s_4=0)\,,\qquad \vec{n}=\breve{n}\,.
\end{equation}
and the seed matrix $\Phi$ is now 
\begin{equation}
	\Phi_{\vec{s}}(x,y,0)=e^{  \sinh\gamma y}e^{\Sigma_+x} S+e^{- \sinh\gamma y}e^{\Sigma_-x}\,,\qquad S=\left(	\begin{array}{cc}
		s_1 & s_2+\Ti s_3 \\
		-s_2+\Ti s_3 & s_1 \\
	\end{array}
	\right)\,.
\end{equation}

After these reductions, it is easy to verify that the final seed matrix admits the representation (\ref{condition2}) and, due to the relation (\ref{NonSin}), the resulting DS II soliton will be free of singularities. Indeed, the form of the soliton at $\tau=0$ is given by  
\begin{align}
    &u_1(x,y,0)=	1-V_{\vec{s}} (x,y,0)+\Ti\, m_{\vec{s}}( x,y,0)\,, \\
	\label{paramV0}
	&V_{\vec{s}}(x,y,0)= \frac{ 2    \sinh\gamma[(\sinh\gamma s_1-s_2)  \cos  x_1-(s_1+\sinh \gamma s_2) \sinh\gamma]  }
	{\sinh \gamma  [(\sinh\gamma s_1-s_2)  \cos  x_1]+\cosh ^2\gamma  \cosh x_2+s_1+\sinh \gamma s_2}\,,
	\\
	\label{paramm0}
	&m_{\vec{s}}(x,y,0)= \frac{2  \sinh \gamma  \cosh ^2\gamma s_3}{\sinh \gamma  [(\sinh\gamma s_1-s_2)  \cos x_1]+\cosh ^2\gamma  \cosh x_2+s_1+\sinh \gamma s_2} \,,
\end{align}
where, for the sake of simplicity, we have summoned the abbreviated notation (\ref{notation}).

\begin{figure}[H]
  	\begin{center}
  		\includegraphics[scale=0.5]{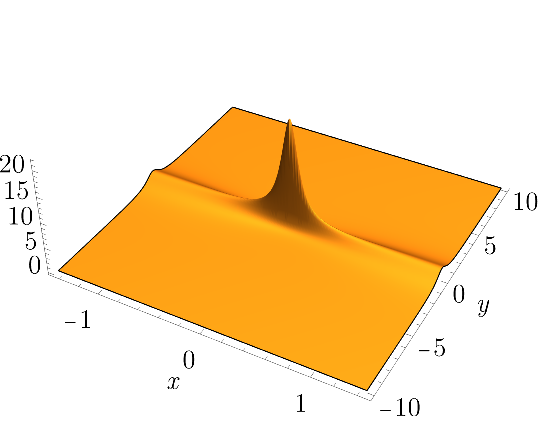}\quad
  		\includegraphics[scale=0.5]{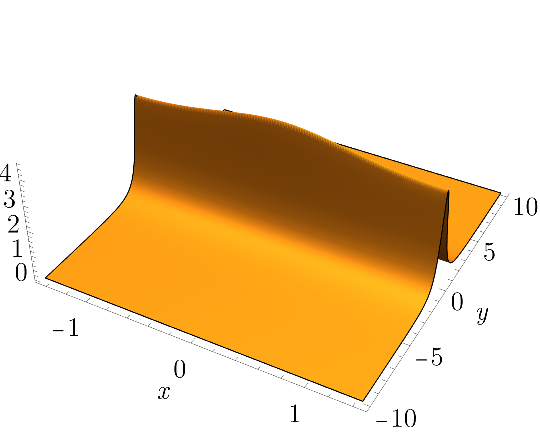}\quad
  		\includegraphics[scale=0.5]{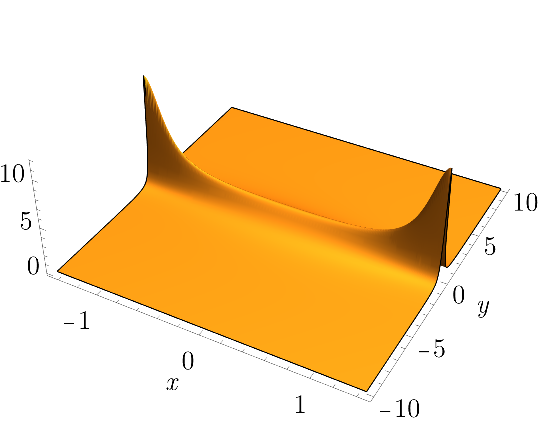}
  	\end{center}
  	\caption{\small Different plots of $V_{\gamma,\varphi}(x,y,0)$.  From left to right  the parameters are $\gamma=0.4$ and  $\varphi=\{\tfrac{9}{10}\pi,\,\tfrac{99}{100}\pi,\tfrac{108}{100}\pi\}$. }
  	\label{figPotFam}
  \end{figure}

The initial condition (\ref{initialcond}) forces us to fix $s_3=0$, see (\ref{paramm0}). This restricts $s_1$ and $s_2$ to be the coordinates of the equatorial circle in Figure \ref{FigDiag}. It is convenient to parametrize 
\begin{equation}
	\label{CircleofSolutions}
	(s_1,s_2)= (\text{sech}\gamma  \cos \varphi -\tanh \gamma  \sin \varphi ,\tanh\gamma  \cos \varphi +\text{sech}\gamma  \sin \varphi)\,,
\end{equation}
such that the two points invariant under $x$-translations are identified with values $\varphi=0$ and $\varphi=\pi$. With this choice, the Hermitian Hamiltonian at $\tau=0$ takes the form 
\begin{align}
\label{H0phi}
&	H_1= -\Ti (  \sigma_1\partial_x+\sigma_2 \partial_y)+V_{\gamma,\varphi}(x,y)=: H_{\gamma,\varphi}(0)\,.
	\\
&	V_{\gamma,\varphi}(x,y,0)=
	\frac{2   \tanh \gamma  (\sinh \gamma  \cos \varphi +\sin \varphi  \cos(2 x\cosh\gamma ))}{\tanh \gamma  \sin \varphi  \cos  (2 x\cosh\gamma)-\text{sech}\gamma  \cos \varphi -\cosh  (2y\sinh\gamma)}\,.
\end{align}
Variations of the parameters $\gamma$ and $\varphi$ can change the shape of the potential, modify the position of the extremes, and introduce saddle points. Some plots for different values of $\varphi$ and $\gamma$ are shown in Fig. \ref{figPotFam}.

  Since at the two stability points the configuration is invariant under $x$-translations, the corresponding potential becomes one-dimensional, i.e., it depends only on $y$, 
    \begin{align}
  	&V_{\gamma,0}(x,y,0)=-
  	\frac{2  \sinh ^2\gamma  }{1  +\cosh\gamma\cosh   (2 y\sinh\gamma)} \,,
  	&V_{\gamma,\frac{\pi}{2}}(x,y,0)=-
  	\frac{2 \sinh ^2\gamma  }{1  -\cosh\gamma\cosh   (2 y\sinh\gamma)} \,.
  \end{align}
The potential in  (\ref{H01}) corresponds to the case $(s_1,s_2)=(1,0)$ or equivalently $\varphi=-\tan ^{-1}(\sinh \gamma )$. 

\begin{figure}[H]
	\begin{center}
		\includegraphics[scale=0.55]{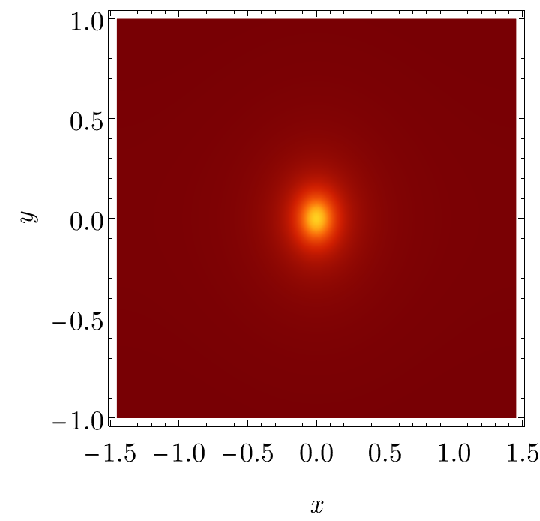}
		\includegraphics[scale=0.55]{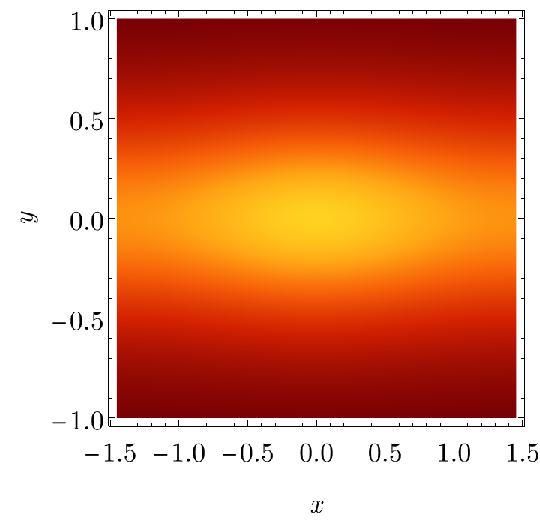}
	     \includegraphics[scale=0.55]{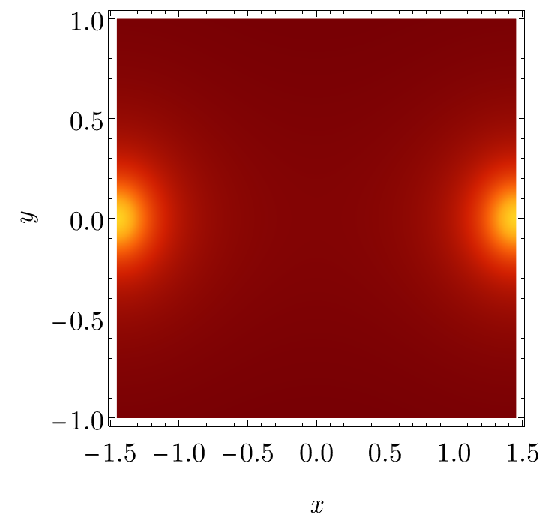}
	\end{center}
	\caption{\small Plots of the probability density. From left to right  the parameters are $\gamma=0.4$ and  $\varphi=\{\tfrac{9}{10}\pi,\,\tfrac{99}{100}\pi,\,\tfrac{108}{100}\pi\}$. }
	\label{figRhophi}
\end{figure}

To conclude this section, we explicitly show the form of $\Sigma=\Sigma_{\gamma,\varphi}(x,y,0)$ in terms of the abbreviated variables (\ref{notation}),
\begin{align}
	& \Sigma_{\gamma,\varphi}(x,y,0)= \frac{A_{\gamma,\varphi}(x,y,0)}
	{ \tanh \gamma  \sin \varphi  \cos x_1-\cosh x_2-\text{sech}\gamma  \cos \varphi}\,,
	\\
	&A_{\gamma,\varphi}(x,y,0)=\left(
	\begin{array}{cc}
		- (\sin \varphi \sin x_1-\Ti \sinh x_2)\sinh \gamma  & \Ti\left(\cosh x_2+\cos \varphi  \cosh \gamma \right) \\
		\Ti\left(\cosh x_2+\cos \varphi  \cosh \gamma\right) & -   (\sin \varphi  \sin x_1+\Ti \sinh x_2)\sinh \gamma \\
	\end{array}
	\right)\,.
\end{align}
It is easy to verify that $\Sigma_{\gamma,\varphi}(x,y,0)\rightarrow \Sigma_\pm$ when $y\rightarrow \pm \infty$, therefore the asymptotic properties are as expected. The SKT and the bound states with energy $E=1$ are obtained by using formulae (\ref{SKTstates0}) and (\ref{bound1})-(\ref{bound4}), with the change 
\begin{equation}
	L_\gamma (\tau)\rightarrow L_{\gamma,\varphi}(0)\,,\qquad 
L_{\gamma,\varphi}(0)=\partial_x-\Sigma_{\gamma,\varphi}(x,y,0)\,.
\end{equation} 
In this regard, one can directly compute the probability density for the bound states, and as a result we get 
\begin{equation}
	\rho_{\gamma,\varphi}(x,y,0)\propto \frac{1}{ \tanh \gamma  \sin \varphi  \cos x_1-\cosh x_2-\text{sech}\gamma  \cos \varphi}\,.
\end{equation}
Fig. \ref{figRhophi} shows that a variation of the parameter $\varphi$ modifies the localization of the particle.
 
\subsection{Three-parametric $\mathcal{PT}$-symmetric family at $\tau\not=0$}
\label{SubSecTnot0}

\begin{figure}[H]
	\begin{center}
		\includegraphics[scale=0.5]{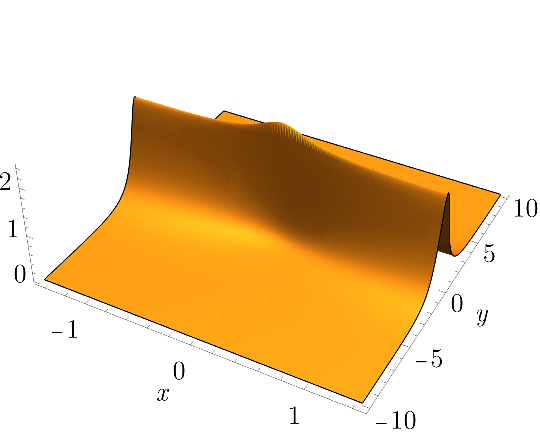}
		\includegraphics[scale=0.5]{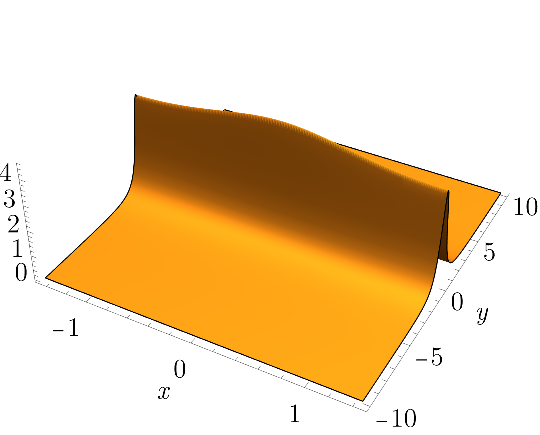}
		\includegraphics[scale=0.5]{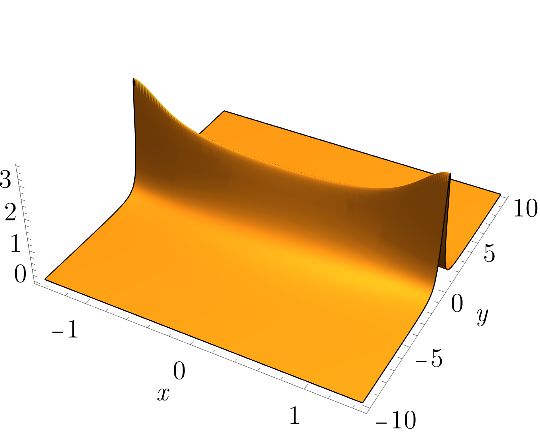}	
	\end{center}
	\caption{\small Plots of potential $V_{\gamma,\varphi}(x,y,\tau)$.
	From left to right  the parameters are $\gamma=0.4$, $\tau=0.12$, and	 $\varphi=\{\tfrac{9}{10}\pi,\,\tfrac{99}{100}\pi,\,\tfrac{108}{100}\pi\}$.  }
	\label{figPotphiTau}
\end{figure}

\begin{figure}[H]
	\begin{center}
		\includegraphics[scale=0.5]{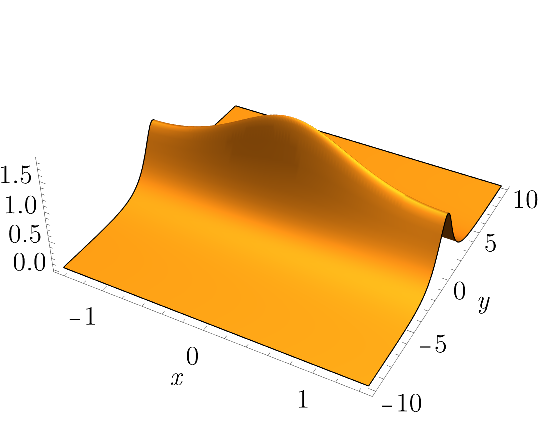}
		\includegraphics[scale=0.5]{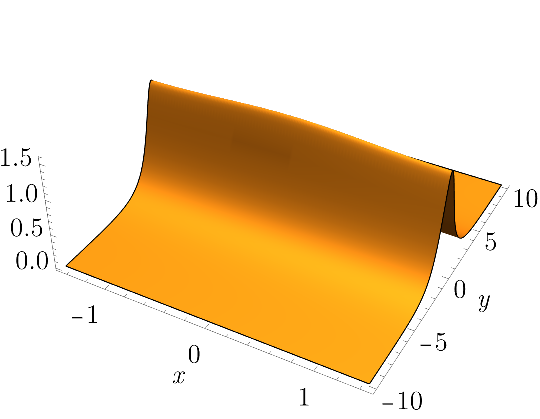}
		\includegraphics[scale=0.5]{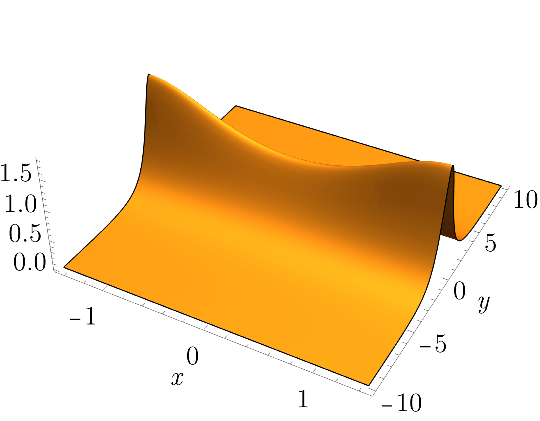}	
	\end{center}
	\caption{\small Plots of mass term $m_{\gamma,\varphi}(x,y,\tau)$.
		From left to right  the parameters are  $\gamma=0.4$, $\tau=0.5$, and	 $\varphi=\{\tfrac{9}{10}\pi,\,\tfrac{99}{100}\pi,\,\tfrac{108}{100}\pi\}$.  }
	\label{figMas2phiTau}
\end{figure}

In order to restore the $\tau$-dependence we let the operator 
$e^{2\tau\partial_x\partial_y}$ act on the general seed matrix. By using (\ref{SimB}), one realizes that this action is equivalent to taking the seed matrix solution as
\begin{align}
	\Phi_{\vec{s}}(x,y,\tau) = e^{\sinh\gamma y}e^{\Sigma_+x}S(\tau)+e^{-\sinh\gamma y}e^{\Sigma_-x}\,,\qquad S(\tau)=e^{2 \tau  \sinh \gamma \Sigma _+ }\,S e^{2 \tau  \sinh \gamma\Sigma _- }\,.
\end{align}
From the second equation one reads that the $\tau-$translation is a rotation  in the $S^2(s_4=0)$ sphere (\ref{ParameterSpace}) along a fixed axis. Its explicit form is ($i=1,2,3$)
\begin{equation}
	\label{tautrans}
	s_i\rightarrow s_i(\tau)=Q_{ij}(\tau)s_j\,,\qquad Q(\tau)=\exp(x_0 \hat{X}_\gamma)\,,\qquad \hat{X}_\gamma=\left(\begin{array}{ccc}
		0 &0 & \sech\gamma \\
		0 & 0 &\tanh\gamma \\
		-\sech\gamma & -\tanh\gamma & 0
	\end{array}\right)\,,
\end{equation}
with $x_0= 2\tau \sinh\gamma$. 
There are two stability points that lie on the equatorial circle $s_3=0$  given by
\begin{equation}
	\vec{s}=\pm \vec{q}\,,\qquad \vec{q}=(-\tanh\gamma,\sech \gamma,0)\,.
\end{equation}
These points correspond to stationary DS II solutions. The entire $S^2(s_4=0)$ sphere is then recovered by applying the transformation (\ref{tautrans}). The points of this sphere are  
\begin{equation}
	\vec{s}(x_0)=(\text{sech}\gamma  \cos \varphi  \cos x_0-\tanh \gamma  \sin \varphi ,\tanh \gamma  \cos \varphi  \cos x_0+\text{sech}\gamma  \sin \varphi ,-\cos \varphi  \sin x_0)\,.
\end{equation}
and it is notable that  $\vec{s}(x_0)$ reduces to the $\tau$-translation invariant point $\pm \vec{q}$ for $\varphi=\pm \tfrac{\pi}{2}$. With this choice, the final form of the DS II solution has the form (\ref{GenericSoliton}), where the functions $V(x,y,\tau)$ and $m(x,y,\tau)$ are given by 
\begin{align}
	\label{paramV}
	&V_{\gamma,\varphi}(x,y,\tau)= \frac{2   \tanh \gamma  \left(\sinh \gamma  \cos x_0 \cos \varphi +\sin \varphi  \cos x_1\right)}{ \tanh \gamma  \sin \varphi  \cos x_1-\cosh x_2-\text{sech}\gamma  \cos x_0 \cos \varphi}\,,
	\\
	&m_{\gamma,\varphi}(x,y,\tau)=-\frac{2  \sinh \gamma  \sin x_0 \cos \varphi }{\tanh \gamma  \sin \varphi  \cos x_1- \cosh x_2-\sech\gamma\cos x_0 \cos \varphi }\,.\label{paramm}
\end{align}
Then, the resulting Dirac Hamiltonian corresponds to 
\begin{equation}
	\label{Htau}
	H_1=-\Ti(\sigma_1\partial_{x}+\sigma_2\partial_y)+V_{\gamma,\varphi}(x,y,\tau)+\Ti \, m_{\gamma,\varphi} (x,y,\tau)\sigma_3=: H_{\gamma,\varphi}(\tau)\,.
\end{equation}
Some plots of the potential and the mass term for different values of $\tau$ are shown in Fig. \ref{figPotphiTau} and Fig. \ref{figMas2phiTau}, respectively. 

The system is invariant under the $\mathcal{T}$ and $\mathcal{P}_x\mathcal{P}_y$ transformation and share identical properties of the system discussed in Sec. \ref{SecExp}. At the $\tau$-translation invariant point $\varphi=\pm \frac{\pi}{2}$, the corresponding Hamiltonian is Hermitian, independent of $\tau$, and takes the form 
\begin{equation}
	\label{Htau}
	H_{\gamma,\pm\frac{\pi}{2}}(\tau)=-\Ti(\sigma_1\partial_{x}+\sigma_2\partial_y)\mp  \frac{2\tanh \gamma  \cos  (2x \cosh\gamma)}{\cosh  (2y \sinh\gamma)\pm \tanh \gamma  \cos (2x\cosh\gamma)}\,.
\end{equation}

\begin{figure}[H]
	\begin{center}
		\includegraphics[scale=0.5]{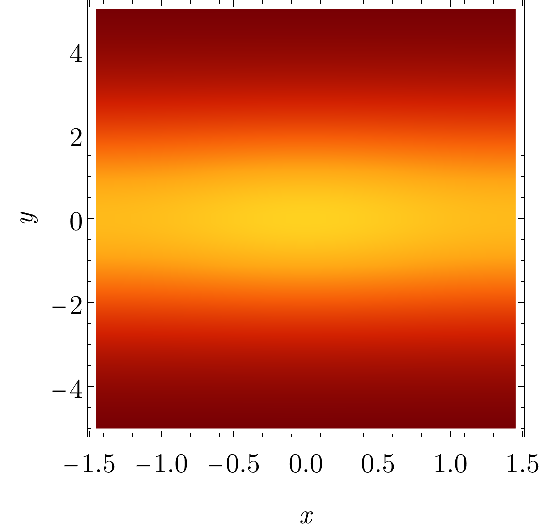}
		\includegraphics[scale=0.5]{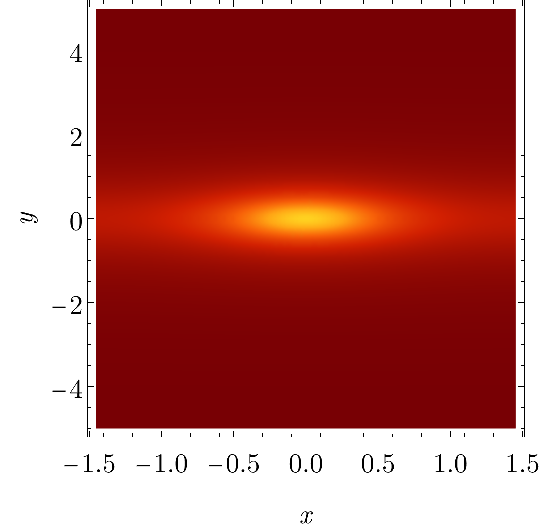}
		\includegraphics[scale=0.5]{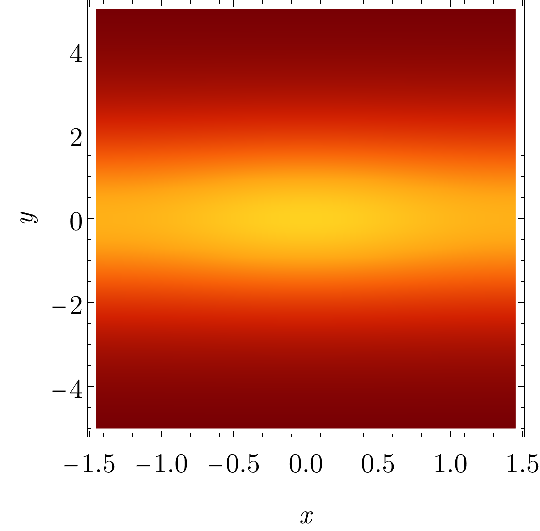}	
	\end{center}
	\caption{\small Plots of the probability density with parameters  $\gamma=0.4$, $\varphi=\tfrac{9}{10} \pi$ and $\tau=-1.5,0.1,1.5$ (from left to right). Changes in  $\tau$ affect the dispersion. }
	\label{figRhoTau}
\end{figure}

Similar to the case $\tau=0$ case, the SKT and bound state solutions with energy $E=1$ are computed by following formulas 
(\ref{SKTstates0}) and (\ref{bound1})-(\ref{bound4}), but now with the change  
\begin{equation}
	L_\gamma (\tau)\rightarrow L_{\gamma,\varphi}(\tau)\,,\qquad 
	L_{\gamma,\varphi}(\tau)=\partial_x-\Sigma_{\gamma,\varphi}(x,y,\tau)\,,
\end{equation} 
where \begin{align}
	& \Sigma_{\gamma,\varphi}(x,y,\tau)= \frac{ A_{\gamma,\varphi}(x,y,\tau)}
	{ \tanh \gamma  \sin \varphi  \cos x_1-\cosh x_2-\text{sech}\gamma  \cos x_0 \cos \varphi}\,,
	\\
	&A_{\gamma,\varphi}(x,y,\tau)=\left(
	\begin{array}{cc}
		- (\sin \varphi \sin x_1-\Ti \sinh x_2)\sinh \gamma  & \Ti\left(\cosh x_2+\cos \varphi  \cosh \left(\gamma -\Ti x_0\right)\right) \\
		\Ti\left(\cosh x_2+\cos \varphi  \cosh (\gamma +\Ti x_0)\right) & -   (\sin \varphi  \sin x_1+\Ti \sinh x_2)\sinh \gamma \\
	\end{array}
	\right)\,.
\end{align}
The probability density associated with the four bound states is given by 
\begin{equation}
	\rho_{\gamma.\varphi}(x,y,\tau)\propto \frac{1}{ \tanh \gamma  \sin \varphi  \cos x_1-\cosh x_2-\text{sech}\gamma  \cos x_0 \cos \varphi}\,.
\end{equation}
Plot of this function for different values of $\tau$ are shown in Fig. \ref{figRhoTau}. 

\subsection{Three-parametric Hermitian family for $\tau=\Ti z$}
\label{SubSecTim}
\begin{figure}[H]
	\begin{center}
		\includegraphics[scale=0.5]{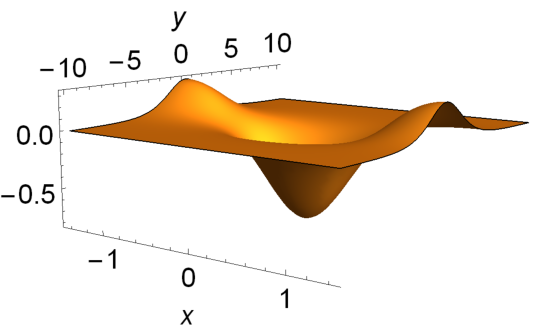}\quad
		\includegraphics[scale=0.5]{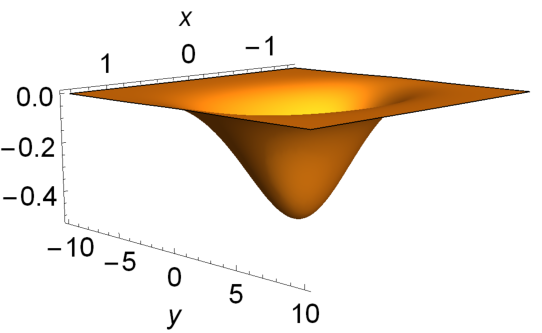}\quad
		\includegraphics[scale=0.5]{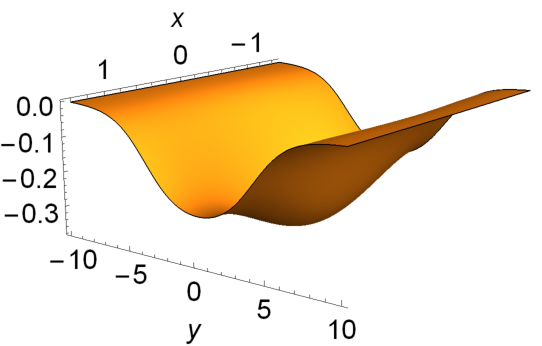}	
	\end{center}
	\caption{\small Plots of potential $\widetilde{V}_{\gamma,\varphi}(x,y,z)$ with parameters $\gamma=0.4$, $\varphi=1.25$ and $z=0.5\,,1.5\,,2.5,$ (from left to right).   }
	\label{figPotphiTauz}
\end{figure}

\begin{figure}[H]
	\begin{center}
		\includegraphics[scale=0.5]{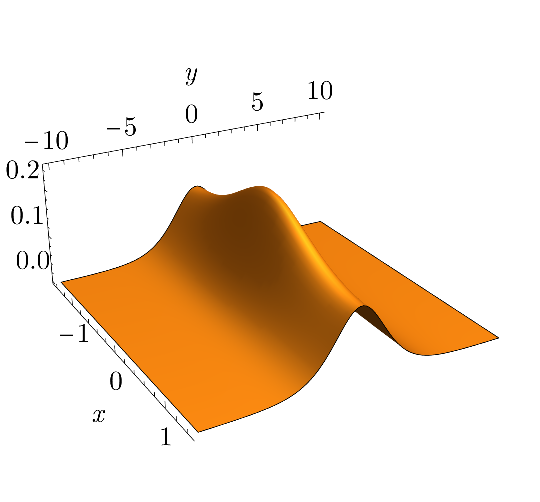}\,
		\includegraphics[scale=0.5]{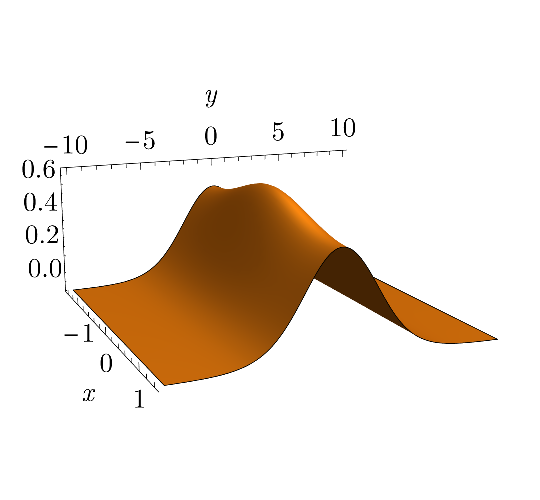}\,
		\includegraphics[scale=0.5]{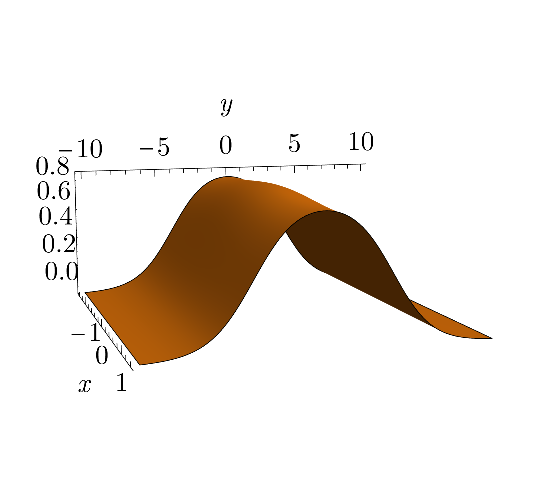}	
	\end{center}
	\caption{\small Plots of mass term $\widetilde{m}_{\gamma,\varphi}(x,y,\tau)$. The parameters $\gamma=0.4$, $\varphi=1.25$ and $z=0.5\,,1.5\,,2.5,$ (from left to right).  }
	\label{figMasz$phiTau}
\end{figure}

When $\tau$ is imaginary, we obtain a parametric Hermitian but no longer $\mathcal{T}$ symmetric  Dirac Hamiltonian
\begin{align}
	\label{Htau}
 &H_{\gamma,\varphi}(\Ti z)=-\Ti(\sigma_1\partial_{x}+\sigma_2\partial_y)+\widetilde{V}_{\gamma,\varphi}(x,y,z)+\, \widetilde{m}_{\gamma,\varphi} (x,y,z)\sigma_3=: \widetilde{H}_{\gamma,\varphi}(z)\,,
\\
	\label{paramV}
	&\widetilde{V}_{\gamma,\varphi}(x,y,z)= \frac{2  \tanh \gamma  \left(\sinh \gamma  \cosh x_3 \cos \varphi +\sin \varphi  \cos x_1\right)}{ \tanh \gamma  \sin \varphi  \cos x_1-\cosh x_2-\text{sech}\gamma  \cosh x_3 \cos \varphi}\,,
	\\
	&\widetilde{m}_{\gamma,\varphi}(x,y,z)=\frac{2   \sinh \gamma  \sin x_3 \cos \varphi }{\tanh \gamma  \sin \varphi  \cos x_1- \cosh x_2-\sech\gamma\cosh x_3 \cos \varphi }\,,\label{paramm}
\end{align}
where $x_3=2  z \sinh2\gamma$ as given in (\ref{x3}). Plots of the potential and mass term are shown in Figs. \ref{figPotphiTauz} and \ref{figMasz$phiTau}, respectively. 

One realizes that the denominator has a zero for a particular subset of the three parameters $\gamma$, $\varphi$ and $z$. This can also be seen from the seed solution, which reads  
\begin{align}
	&	\Phi_{\vec{s}} (x,y,\Ti z) = e^{\cosh\gamma y}e^{\Sigma_+x}G+e^{-\cosh\gamma y}e^{\Sigma_-x}=: {\Phi}_{\vec{g}}(x,y,z)\,,
	\\
	& G= \left(
	\begin{array}{cc}
		g_1 & g_2+g_3\\
		-g_2+ g_3 & g_1 \\
	\end{array}
	\right)\,,\qquad g_1^2+g_2^2-g_3^2=1\,, 
\end{align}
where, in contrast to the  two-sphere in the $\mathcal{PT}$-symmetric case,  $\vec{g}=(g_1,g_2,g_3)$ denotes a point on a one sheet hyperboloid with coordinates 
\begin{equation}
	\vec{g}= (\text{sech}\gamma  \cos \varphi  \cosh x_3-\tanh \gamma  \sin \varphi ,\tanh \gamma  \cos \varphi  \cosh x_3+\text{sech}\gamma  \sin \varphi ,-\cos \varphi  \sinh x_3)\,.
\end{equation}
The matrix ${\Phi}_{\vec{g}}(x,y,z)$ is not invariant under the action of $\mathcal{T}$, and therefore the relation (\ref{NonSin}) does not longer hold, i.e., the denominator of the potential is not necessary positive. To prevent poles in the system we must demand 
\begin{equation}
 \tanh \gamma  \sin \varphi -\sech\gamma \cos \varphi \cosh x_3<1 \,.
\end{equation}

The transformation $\mathcal{T}$ flips between $g_3>0$ and $g_3<0$ on the hyperbolic parameter space, i.e., 
\begin{equation}
	\mathcal{T}{\Phi}_{\vec{g}}(x,y,z)={\Phi}_{\vec{g}}(x,y,-z)	\mathcal{T}\,.
\end{equation}  
This behaviour is also inherited by the Hamiltonian and the intertwining operator,
\begin{equation}
	\mathcal{T}\widetilde{H}_{\gamma,\varphi}(z)=\widetilde{H}_{\gamma,\varphi}(-z)	\mathcal{T}\,,\qquad 
	\mathcal{T}\widetilde{L}_{\gamma,\varphi}(z)=\widetilde{L}_{\gamma,\varphi}(-z)	\mathcal{T}\,,\qquad \widetilde{L}_{\gamma,\varphi}(z):=L_{\gamma,\varphi}(\Ti z)
	\,.
\end{equation}

\begin{figure}[H]
 	\begin{center}
 		\includegraphics[scale=0.55]{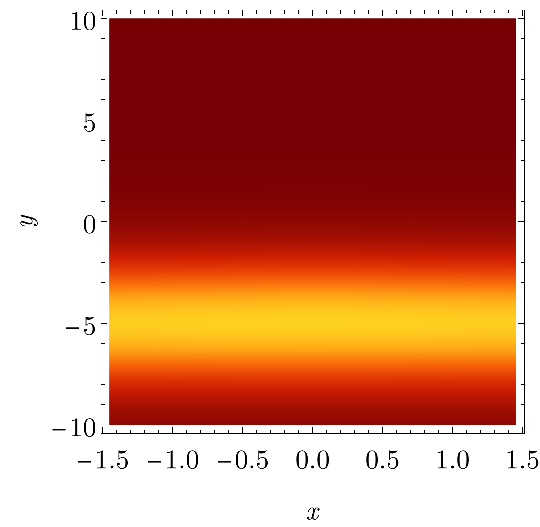}
 		\includegraphics[scale=0.55]{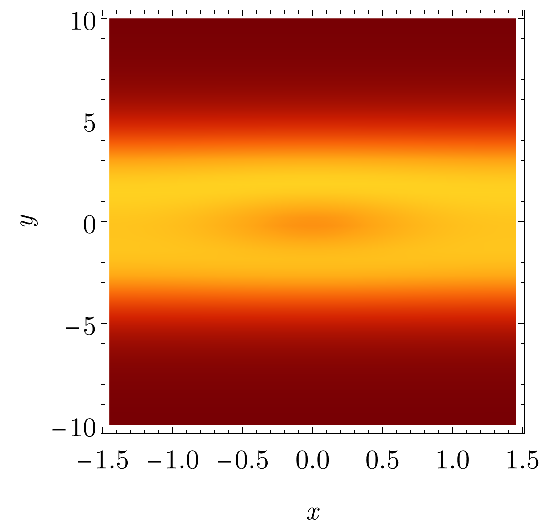}
 		\includegraphics[scale=0.55]{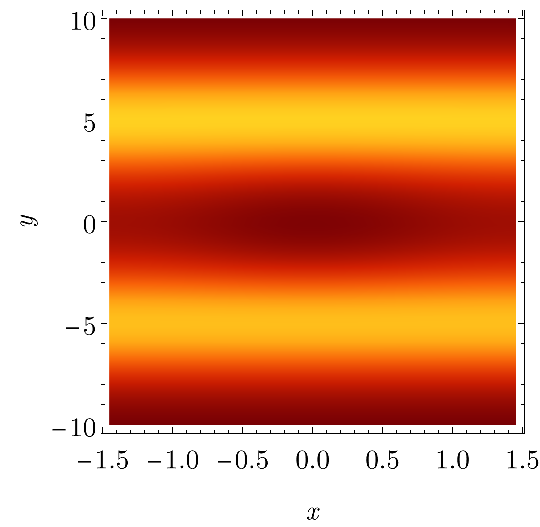}	
 	\end{center}
 	\caption{\small Plots of the bound states probability density. The parameters $\gamma=0.4$, $\varphi=1.25$ and $z=-3\,,1.5\,,3,$ (from left to right). }
 	\label{figBoundphiTau}
 \end{figure}

The SKT states are constructed as usual, but for the bound state we 
use the formulas  (\ref{boundHer1})-(\ref{boundHer4}) with the replacement 
\begin{equation}
	\widetilde{L}_\gamma (z)\rightarrow 	\widetilde{L}_{\gamma,\varphi}(z)\,. 
\end{equation}
As in the example, we have four bound states with two different probability densities, corresponding to $\rho_{\gamma,\varphi}(x,y,\pm z)$. We do not write the full expressions; instead, we show density plots for different parameter values in Fig. \ref{figBoundphiTau}.

\section{Underlying supersymmetry }

\label{SecSUSY}

In nonrelativistic one-dimensional quantum mechanics, reflectionless systems can be understood as supersymmetric partners of the free particle. This structure makes it possible to construct higher-order analogs of the momentum operator through ``Darboux dressing''. The eigenvalues of such operators are directly related to the wave numbers of the scattering states, and they usually annihilate all isolated bound states in the spectrum. In this section, we demonstrate that an analogous structure emerges for the Dirac systems constructed in the previous section. In this case, however, the corresponding Darboux-dressed operators are not integrals of motion; instead, they act as quasi-symmetry generators, in the sense introduced in \cite{Quasi}.

Since the construction is completely general and applies to arbitrary values of the parameters $\gamma$ and $\varphi$, we will omit the corresponding indices. Likewise, we suppress the explicit dependence on $\tau$, indicating only when its real or imaginary character is relevant. Finally, for the manipulation of states, it is convenient to introduce the Dirac braket notation,
\begin{equation}
	\ket{\chi}=L\ket{\theta}\,,\qquad \braket{\vr}{\theta}=\psi(x,y|\theta)\,.
\end{equation}
where $\psi(x,y|\theta)$ is given by (\ref{FreeSpinors}), so 
\begin{equation}
	H_0\ket{\theta}=-\ket{\theta} \,,\qquad H_1\ket{\chi}=\ket{\chi}\,.
\end{equation}
A familiar starting point to establish a supersymmetric relation between two quantum Hamiltonians is the existence of an intertwining relation. In our case, we only have the asymmetric version (\ref{asym})
\begin{equation}
	\label{Lop}
	\sigma_2L \sigma_2 (H_0+1)=(H_1-1)L\,,\qquad L=\partial_x-\Sigma.
\end{equation}
By Hermitian conjugation (in the $\mathcal{PT}$-symmetric case with real $\tau$, we also need to apply $\tau\rightarrow -\tau$) we obtain another relation of this nature  
\begin{equation}
\label{Yop}
	\sigma_2Y \sigma_2 (H_1-1)=(H_0+1)Y\,,\qquad 
	Y=-\partial_x-\Sigma\,.
\end{equation}
The new intertwining operator $Y$ maps the eigenstates of the transformed system $H_1$ with energy $E=1$ into a free-particle solution with energy $E=-1$, i.e., it is a reverse transformation. In fact, by calculating the action of this operator, we obtain
\begin{equation}
	\label{FreeNonRe}
Y\ket{\chi}=YL\ket{\theta}=\mathcal{H}_0\ket{\theta}=(\cos^2\theta-\cosh^2\gamma)\ket{\theta}\,,
\end{equation}
where $\mathcal{H}_0$ 
\begin{equation}
\mathcal{H}_0=YL=-\partial_x^2-\cosh\gamma^2\,,
\end{equation}
is a nonrelativistic one-dimensional free-particle Hamiltonian. Equation (\ref{FreeNonRe}) indicates that the operator $Y$ annihilates the bound states of the system $H_1$ ($\theta=\pm \Ti \gamma$), which is consistent with the absence of bound states for the free particle. 

On the other hand, we can also introduce the reversed product or superpartner
\begin{equation}
\mathcal{H}_1:=L\,Y=
\mathcal{H}_0-2\partial_x\Sigma\,.
\end{equation}
With this object we can build a standard  intertwining relation 
\begin{equation}
	L\,\mathcal{H}_0=\mathcal{H}_1\,L \,,\qquad
	Y\,\mathcal{H}_1=\mathcal{H}_0\,Y \,,
\end{equation}
which allows us to compute the action of $\mathcal{H}_1$ on the transformed states, 
\begin{equation}
\mathcal{H}_1\ket{\chi}=L\,Y\,L\ket{\theta}=L\,\mathcal{H}_0\ket{\theta}=(\cos^2\theta-\cosh^2\gamma)\ket{\chi}\,.
\end{equation}
We learn that $\mathcal{H}_1$ operates as a symmetry transformation at least in the subspaces $E=1$ of states of our Dirac system $H_1$. Moreover, the operator reveals information about the incident angle $\theta$ and, by annihilation, detects the set of bound states. Surprisingly, as we will see below, $\mathcal{H}_1$ does not commute with the Dirac Hamiltonian $H_1$ and therefore it is not a symmetry of the entire system. Rather, it is a quasi-symmetry enjoyed by a single and unique energy level. 

The existence of the operator $Y$ also enables the construction of a family of quasi-symmetry operators, defined as follows:
\begin{equation}
\label{OpO}
	\mathcal{I}=L\,I\,Y,
\end{equation}
where $I$ is an arbitrary symmetry operator of the massless free Dirac particle, i.e., $[H_0,I]=0$, for example any element of the Euclidean group. In fact, we have on the one hand from (\ref{Lop}) and (\ref{Yop}) that 
\begin{equation}
[H_1,\mathcal{I}]=[H_1-1,\mathcal{I}]=(\sigma_2L\sigma_2\,I\,\sigma_2 Y\sigma_2-L\,I\,Y)(H_1-1)\,,
\end{equation} 
which is in general nonzero. On the other hand, from the last factor on the right-hand side of this equation we learn that $[H_1,\mathcal{I}]\ket{\chi}=0$ implies
\begin{equation}
H_1\,\mathcal{I}\ket{ \chi}=\mathcal{I} H_1\,\ket{\chi}= \mathcal{I}\,\ket{\chi}\,,
\end{equation}
so $\mathcal{I}\ket{\chi}$ is another eigenstate of our Dirac system. We can therefore conclude that the subspace of SKT states exhibits several features characteristic of a free particle. The quasi-symmetry transformations mirror the symmetries of a massless free particle, which survive only at this specific energy level. Indeed, up to an overall constant factor, these transformations act solely on the plane-wave component,
\begin{equation}
\mathcal{I}\ket{\chi}=L\,I\,Y\,L\ket{\theta}=(\cos^2\theta-\cosh^2\gamma)L\,I\ket{\theta}\,.
\end{equation}
 All bound states at this energy level are annihilated by these transformations due to the presence of $Y$ in the definition (\ref{OpO}).
Since $I$ can be identified with spatial rotations, or more generally with any generator of the Euclidean group, the infinite set of states with $E=1$ for $H_1$ can be generated by Darboux-dressed rotations,
 \begin{equation}
\mathcal{R}(\alpha)=L e^{\Ti \alpha J} Y\,,\qquad  J=-\Ti (x\partial_y-y\partial_x)+\tfrac{1}{2}\sigma_3\,,\qquad 	\mathcal{R}(\alpha)\ket{ \chi}\propto  L\ket{\theta+\alpha}\,.
 \end{equation}
 
\section{Discussion and outlook}
\label{SecDis}

The main result of this article is the construction of planar Dirac systems that exhibit total omnidirectional transmission, also known as the SKT effect, while simultaneously supporting bound states in the continuum. This is achieved by exploiting the integrability of the nonlinear DS~II model, a $(2{+}1)$-dimensional integrable system that can be reformulated as a pair of linear, matrix-valued partial differential equations. One of these equations can be mapped onto a stationary Dirac equation at a fixed energy. Within this framework, the scalar potential is determined by the real part of a DS~II solution, the mass term arises from its imaginary part, and the corresponding energy level is fixed by the asymptotic behavior of the solution.

In constructing DS II solutions, we exploit the covariance of the matrix linear formulation under Darboux transformations \cite{Matveev}, which provides a systematic method for generating new solutions from a known one. In our first example, starting from the simplest constant DS II solution, the construction of a Dirac equation exhibiting the SKT effect requires the presence of a breather solution of the integrable system, i.e., a configuration localized on the surface of an infinite cylinder. This initial insight into the nature of the relevant soliton is subsequently confirmed for more general solutions by imposing the SKT effect as a guiding requirement in the construction of admissible DS II solutions and by solving the corresponding asymptotic inverse problem.

This result stands in sharp contrast to the corresponding nonrelativistic one-dimensional problem, where reflectionless models are associated with decaying soliton solutions of the KdV equation. Another important difference is that, within the SKT/DS~II correspondence, the full spectrum of the resulting Dirac system cannot be reconstructed. Instead, explicit solutions can be obtained only at a single energy level. This limitation arises because the Dirac model we construct is related to the free particle through an asymmetric Darboux transformation \cite{asym1,CorInzJak1,CorInzJak2}; see Eqs. (\ref{Lop}) and (\ref{Yop}). Despite this restriction, the resulting subspace of solutions is particularly remarkable, as it supports a hidden supersymmetric structure that is absent at other energy levels. In particular, the corresponding states diagonalize a nontrivial matrix-valued one-dimensional Schr\"odinger Hamiltonian, which is itself a supersymmetric partner of the nonrelativistic free particle. This special relationship allows us to construct Darboux-dressed operators originating from the symmetries of the free particle. These operators preserve the SKT subspace but do not act as genuine symmetries of the full Hamiltonian. In this sense, they constitute quasi-symmetry transformations \cite{Quasi} that are closely related to the Euclidean group. The presence of these quasi-symmetries explains the existence of infinitely many solutions at a single energy level, in contrast to the usual quasi-exactly solvable scenario, where one encounters finite-dimensional and discrete families of solutions \cite{Turbiner1}.

An explicit and irreducible three-parameter family of Dirac Hamiltonians exhibiting the SKT effect has been constructed using this integrability-based approach. Within this family, the scattering states of all systems share a common asymptotic behavior, while differences manifest themselves at the level of the bound states. When the real soliton evolution time is included, the parameter space acquires the topology of an $S^2$ sphere. The corresponding Dirac systems are $\mathcal{PT}$-symmetric and involve an imaginary mass term. Discrete symmetry transformations relate all bound states, leading to identical probability densities whose maxima coincide with the minima of the potential. This feature is absent when the soliton time is taken to be imaginary. In that case, the parameter space instead becomes a one-sheeted hyperboloid, and all resulting systems are Hermitian with a real mass term. The bound states are no longer connected by discrete symmetries, and the corresponding particles localize in one of two distinct spatial regions. The construction of solutions involving decaying wave packets remains an open direction. While examples of Dirac systems with wave-packet potentials have been developed for one-dimensional models \cite{Contreras-Astorga:2019akt}, it would be interesting to explore the physical implications of extending such constructions to two-dimensional systems.

The Darboux transformation is well known to be iterative \cite{Matveev}. Consequently, the approach developed in this article for determining the SKT conditions can be extended to higher-order iterations. It will be interesting to investigate whether the requirement of a breather solution can be relaxed in subsequent transformations. Moreover, new models with qualitatively different features may be obtained by starting from other classes of soliton solutions; see, for example, Refs. \cite{Breather,Jafarian,DS2Sol1,DS2Sol2}. Examples of radially decaying potentials have already been constructed through asymmetric Darboux transformations \cite{CorInzJak1,CorInzJak2}.

Furthermore, one may ask whether the ideas presented here are applicable to other $(2{+}1)$-dimensional integrable models, such as the Kadomtsev--Petviashvili \cite{KP} or Novikov--Veselov systems \cite{NV}, and whether analogous constructions can be pursued for higher-order matrix generalizations \cite{MatrixDSII,MatrixKP}. Another open challenge concerns the determination of solutions at generic energy values. In the present construction, the interactions are separable only in the one-dimensional case, and the associated eigenvalue problems are highly nontrivial. Nevertheless, numerical approaches may provide valuable insight into this problem. In particular, it would be interesting to determine whether additional energy levels can support quasi-symmetry structures.

Finally, it is natural to expect that the present results may be relevant for Dirac materials \cite{DiracMat}. In such systems, effective scalar and mass-like interactions can already be engineered and spatially controlled using available experimental techniques, including chemical doping \cite{Chemi} and mechanical deformation or strain engineering \cite{Pablo}. Since the SKT effect corresponds to robust omnidirectional transparency that is insensitive to the angle of incidence, its realization could provide new mechanisms for waveguiding, filtering, or transport control in mesoscopic electronic devices \cite{GraphElec}. From this perspective, the integrability-based constructions presented here offer a systematic framework for designing transparent Dirac systems with tunable bound-state structures.

\section*{Acknowledgement}
F. C. was supported by Fondecyt Grants No. 1211356 and No. 1252036 and by USACH project 042531CS$\_$Ayudante. L. I. was supported by USACH project 042531CS$\_$Ayudante  and by a Riemann fellowship for visiting researchers at Leibniz Universit\"at Hannover.
L. Inzunza also thanks the Institute of Theoretical Physics at Leibniz Universit\"at Hannover for its hospitality. Additionally, we thank the anonymous referee for the valuable comments.

\end{document}